\documentclass[pdflatex,sn-mathphys-num,oneside]{sn-jnl}

\usepackage{subcaption}
\usepackage{graphicx}%
\usepackage{multirow}%
\usepackage{amsmath,amssymb,amsfonts}%
\usepackage{amsthm}%
\usepackage{mathrsfs}%
\usepackage[title]{appendix}%
\usepackage{xcolor}%
\usepackage{textcomp}%
\usepackage{manyfoot}%
\usepackage{algorithm}%
\usepackage{algorithmicx}%
\usepackage{algpseudocode}%
\usepackage{listings}%
\usepackage{tikz}%
\usepackage{threeparttable}
\usepackage{float}
\usepackage{booktabs}
\usetikzlibrary{shapes.geometric, arrows}%
\tikzstyle{startstop} = [rectangle, rounded corners, 
minimum width=3cm, 
minimum height=1cm,
text centered, 
draw=black, 
fill=red!30]

\tikzstyle{io} = [trapezium, 
trapezium stretches=true, % A later addition
trapezium left angle=70, 
trapezium right angle=110, 
minimum width=3cm, 
minimum height=1cm, text centered, 
draw=black, fill=blue!30]

\tikzstyle{process} = [rectangle, 
minimum width=3cm, 
minimum height=1cm, 
text centered, 
text width=3cm, 
draw=black, 
fill=orange!30]

\tikzstyle{decision} = [diamond, 
minimum width=3cm, 
minimum height=1cm, 
text centered, 
draw=black, 
fill=green!30]
\tikzstyle{arrow} = [thick,->,>=stealth]
\theoremstyle{thmstyleone}%
\theoremstyle{thmstyletwo}%

\theoremstyle{thmstylethree}%

\begin{document}

\title[Article Title]{Physics-Informed GCN-LSTM Framework for Long-Term Forecasting of 2D and 3D Microstructure Evolution}


\author*{\fnm{Hamidreza} \sur{Razavi}}\email{hamidreza.razavi@kuleuven.be}

\author{\fnm{Nele} \sur{Moelans}}\email{nele.moelans@kuleuven.be}

\affil*{\orgdiv{Department of Materials Engineering}, \orgname{KU Leuven}, \orgaddress{\street{Kasteelpark Arenberg 44 Bus 2450}, \city{Leuven}, \postcode{3001}, \country{Belgium}}}

%\affil[2]{\orgdiv{Department}, \orgname{Organization}, \orgaddress{\street{Street}, \city{City}, \postcode{10587}, \state{State}, \country{Country}}}

%\affil[3]{\orgdiv{Department}, \orgname{Organization}, \orgaddress{\street{Street}, \city{City}, \postcode{610101}, \state{State}, \country{Country}}}


\abstract{This paper presents a physics-informed framework that integrates graph convolutional networks (GCN) with long short-term memory (LSTM) architecture to forecast microstructure evolution over long time horizons in both 2D and 3D with remarkable performance across varied metrics. The proposed framework is composition-aware, trained jointly on datasets with different compositions, and operates in latent graph space, which enables the model to capture compositions and morphological dynamics while remaining computationally efficient. Compressing and encoding phase-field simulation data with convolutional autoencoders and operating in Latent graph space facilitates efficient modeling of microstructural evolution across composition, dimensions, and long-term horizons. The framework is capable of capturing the spatial and temporal patterns in evolving microstructures, making it suitable for learning their dynamics. The framework captures the spatial and temporal patterns of evolving microstructures while enabling long-range forecasting at reduced computational cost after training.}

\maketitle

\section{Introduction}\label{sec1}

The macroscopic physical and mechanical properties of the materials are highly dependent on the microstructural grains and domains, which differ in their structure, orientation, and chemical composition. Therefore, gaining insight into the formation and evolution of microstructure mechanisms is of utmost importance. Microstructures are inherently complex and thermodynamically unstable, and understanding their dynamics requires extensive theoretical and experimental research. A powerful method within this domain is the phase-field method for simulating microstructural evolution, which is based on a diffused interface description. This diffuse-based interface approach creates a key advantage for the phase-field by eliminating the need to track interfaces during microstructural evolution \cite{moelans2008introduction}. Despite the versatility of the phase-field method in simulating numerous materials processing domains, it faces a significant drawback due to its high computational cost \cite{chafia2024massively}. 

To address the drawback of the high computational cost, numerous machine learning methods have been developed. Some examples of such methods include the multi-generational convolutional-LSTM framework \cite{subedi2025foretelling}, GrainGNN \cite{qin2024graingnn}, convolutional recurrent neural networks \cite{tiwari2025time}, Unet-based artificial neural networks \cite{peivaste2022machine}, PCA coupled with LSTM \cite{montes2021accelerating}, physics-informed neural networks \cite{qiu2022physics}, and recurrent neural networks \cite{hu2022accelerating}. Although these works represent valuable efforts to address the inherent drawbacks of the phase-field approach, many either overlook the underlying physics-based equations governing microstructural evolution or fail to capture long-term forecasting. Addressing these elements is essential for the advancement of approaches that truly complement the phase-field method. 

In this paper, a physics-informed framework is presented that integrates graph convolutional networks with LSTMs to forecast 2D and 3D microstructure evolution generated by phase-field simulations over long time horizons. The datasets are compressed using 2D and 3D convolutional autoencoders, which both reduce dimensionality and enable joint training with the GCN-LSTM-PI model to enhance predictive accuracy. Once trained, the framework performs future forecasting considerably faster than conventional phase-field simulations. The proposed model is composition-aware, trained jointly on datasets with different compositions, and is capable of forecasting microstructural evolution of different compositions. 

%\section{Literature Review}\label{sec2}

\section{Method}\label{sec3}

\subsection{Phase Diagram Construction}\label{subsec1}

To identify the appropriate thermodynamic conditions for spinodal decomposition, a binary phase diagram of the Bi-Sb alloy is constructed using the Thermo-Calc software and its COST 531 \cite{kroupa2007development} database. The CALPHAD (Calculation of Phase Diagrams) methodology is the basis of the method used to predict equilibrium phase boundaries based on thermodynamic models. 

Based on the phase diagram generated with the Thermo-Calc software, some of the favorable conditions for spinodal decomposition take place at a temperature level of 350 K (100 degrees Celsius) and a molar fraction of Sb within the range of 0.4 and 0.6. These composition and temperature conditions lie within the miscibility gap, where the system spontaneously undergoes phase separation. Under this condition, the curvature of the Gibbs free energy is negative, and the system is unstable to even small changes in temperature that could trigger spontaneous phase separation \cite{luan2023spinodal}. 

\begin{figure}[H]
\centering
\includegraphics[width=0.8\linewidth]{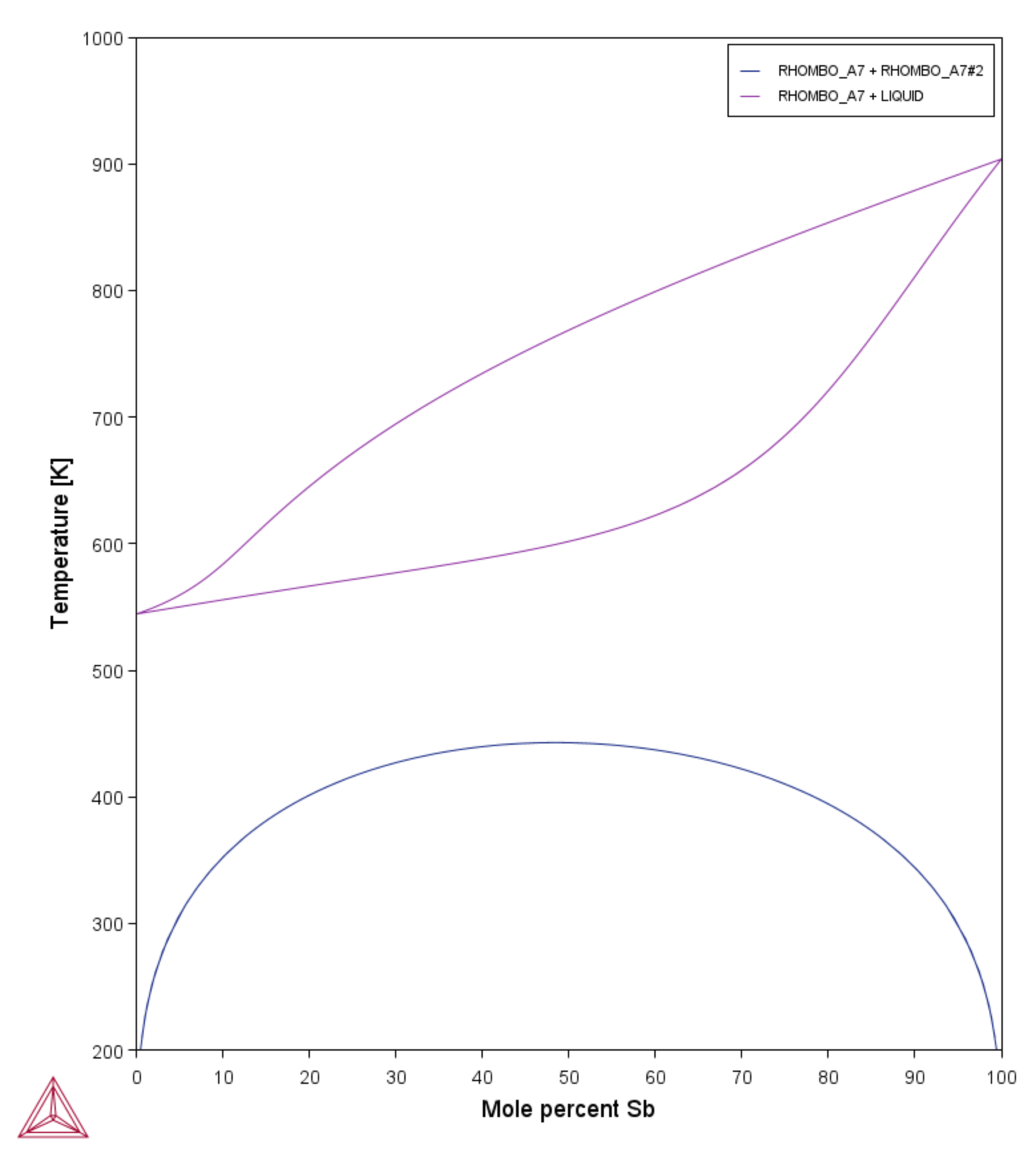}
\caption{Bi--Sb binary phase diagram generated with Thermo-Calc (COST 531 database) using CALPHAD methodology. Spinodal decomposition is predicted to occur near \(T = 350\,\mathrm{K}\) for Sb mole percentage of 40 to 60, conditions that fall within the miscibility gap where the Gibbs free energy curvature is negative and spontaneous separation is thermodynamically favored.}
\label{fig:phase-diagram}
\end{figure}

\subsection{Phase Field Simulation}\label{subsec2}

The phase-field method can simulate the evolution of microstructures in various processes, including spinodal decomposition, phase transformation, grain growth, and solidification \cite{sigala2022phase} \cite{yamanaka2023phase} \cite{moelans2008quantitative} \cite{nomoto2022multi}. This technique employs a diffuse interface approach to model complex morphologies without relying on shape assumptions. Instead of tracking interfaces, it utilizes continuous field variables, known as order parameters, to represent distinct phases or structural states. The order parameters evolve to minimize the total Helmholtz free energy, thereby reaching the equilibrium state that is thermodynamically favored by the system. 

\begin{equation}
F[X] = \int_{\Omega} \left( f(X, T) + \frac{\kappa}{2} |\nabla X|^2 \right) \, d\mathbf{r}
\end{equation}

In the above total Helmholtz free energy function, $f(c)$ is the local free energy density, $\kappa$ is the gradient energy coefficient, and $\nabla $ is the composition gradient. This study focuses on modeling the spinodal decomposition of a Bismuth-Antimony alloy.  For the context of spinodal decomposition, the phase field method is tailored around the Cahn-Hilliard equation. In this approach, the process is governed by a conserved order parameter, which models the spontaneous separation of a homogeneous mixture into distinct regions of varying composition \cite{cahn1958free}. Furthermore, the system evolves so that its chemical potential is spatially homogeneous. (\href{https://doi.org/10.1063/1.1744102}{https://doi.org/10.1063/1.1744102} ) 

Phase-field simulation generates high-resolution data in the context of images to represent the evolution of the microstructure. In this study, the simulation is used to produce 200 images that present two-dimensional compositional fields in sequential order. Second, the simulation provides an accurate model of the spinodal decomposition in the Bi-Sb alloy under the studied thermodynamic conditions. 

This simulation was performed as a two-dimensional model based on the Cahn-Hilliard equation. Under the Cahn-Hilliard model, the phase separation dynamics is described as a system in which the total composition is conserved. As a result, the evolution of the antimony molar field $X_{\text{Sb}}(x, y, t)$ by its governing equation is depicted as follows:

\begin{equation}
\frac{\partial X_{\text{Sb}}}{\partial t} = \nabla \cdot \left( M \nabla \mu \right)
\end{equation}

In the above equation, M is the atomic mobility with a value equal to $1 \times 10^{-26}$ and $\mu$ is the chemical potential. The chemical potential is calculated based on the total Gibbs free energy of the system, incorporating both local chemical contributions and interfacial energies \cite{chen2002phase}. The chemical potential is then given by:
\begin{equation}
\mu = \frac{\delta F}{\delta X_{\text{Sb}}} = \frac{d f}{d X_{\text{Sb}}} - \kappa \nabla^2 X_{\text{Sb}}
\end{equation}

To assess the bulk free energy,  a CALPHAD-based thermodynamic framework is utilized. This framework includes formulas for the molar Gibbs free energies of pure Bi and Sb, as well as interaction coefficients for the two elements. The chemical component of the free energy is represented using Redlich-Kister polynomials. 

\begin{equation}
f(X_{\text{Sb}}, T) = f_{\text{pure}}(X_{\text{Sb}}, T) + f_{\text{mix, ideal}}(X_{\text{Sb}}) + f_{\text{mix, excess}}(X_{\text{Sb}})
\end{equation} 

Where:

\begin{equation}
f_{\text{pure}}(X_{\text{Sb}}, T) = X_{\text{Sb}} G_{\text{Sb}} + (1 - X_{\text{Sb}}) G_{\text{Bi}}
\end{equation}

\begin{equation}
f_{\text{mix, ideal}}(X_{\text{Sb}}) = RT \left( X_{\text{Sb}} \ln X_{\text{Sb}} + (1 - X_{\text{Sb}}) \ln (1 - X_{\text{Sb}}) \right)
\end{equation}

\begin{equation}
f_{\text{mix, excess}}(X_{\text{Sb}}) = X_{\text{Sb}} (1 - X_{\text{Sb}}) \left( L_0 + L_1 (1 - 2X_{\text{Sb}}) \right)
\end{equation}
Each term in the expression of the free energy contributes to the thermodynamic driving force for phase separation. The term $f_{\text{chem}}(X_{\text{Sb}}, T)$ accounts for the reference energies of the pure elements, $f_{\text{mix}}(X_{\text{Sb}})$ captures the ideal configurational entropy of the solution, and $f_{\text{int}}(X_{\text{Sb}})$ represents non-ideal enthalpic interactions between the Bi and Sb atoms. The total derivative of the full free energy $f(X_{\text{Sb}}, T)$ concerning composition, combined with the gradient penalty $\frac{\kappa}{2} |\nabla X_{\text{Sb}}|^2$, determines the chemical potential $\mu$, which in turn governs the flux in the Cahn-Hilliard equation. The expressions for the Gibbs energies of the pure elements $G_{\text{Sb}}$ and $G_{\text{Bi}}$ are temperature dependent and taken from the COST 531 Thermo-Calc database. The interaction coefficients $L_0$ and $L_1$ are also functions of temperature.

The simulation is implemented in MATLAB using a semi-implicit numerical spectral scheme. The computational domain is a $128\times128$ grid with periodic boundary conditions and a spatial resolution of 10 nanometers. The composition field is initialized with a constant background value perturbed by slight random noise to promote spontaneous decomposition. The spatial derivatives in the Cahn-Hilliard equation are efficiently handled in Fourier space. At each time step, the composition field is updated according to:

\begin{equation}
\tilde{X}_{\text{Sb}}^{\,t + \Delta t} = 
\frac{\tilde{X}_{\text{Sb}}^{\,t} - M \Delta t \, k^2 \tilde{\mu}}
     {1 + M \Delta t \, \kappa k^4}
\end{equation}

Where $\tilde{X}_{\text{Sb}}$ and $\tilde{\mu}$ are the Fourier transforms of the composition and chemical potential fields, respectively, $k$ is the magnitude of the wavevectors, $\Delta t$ is the time step, $M$ is the mobility, and $\kappa$ is the gradient energy coefficient. The time evolution was computed over a defined number of steps, with the time step fixed at 100000 timesteps. Every 500 steps, the simulation saves the composition field as MATLAB and Excel files, along with visualizations of both 2D and 1D cross sections. This simulation approach enables the modeling of spinodal decomposition in the Bi-Sb system, while also producing structured image data suitable for training deep learning models.

\subsection{Machine Learning Model Pipeline}\label{subsec3}
%\subsection{2D Convolutional Autoencoder With Input Dimensions 128×128}\label{subsec3}

\usetikzlibrary{shapes.geometric, arrows.meta, positioning}

% Styles
\tikzstyle{rectbox} = [rectangle, rounded corners, minimum width=7cm, minimum height=1cm,
    text centered, align=center, draw=black]
\tikzstyle{arrow}   = [->,>=stealth,thick]

\begin{center}
\begin{tikzpicture}[node distance=1.2cm]

\node (start)  [rectbox, fill=red!30]    {Data Preparation};
\node (ae)     [rectbox, fill=blue!20, below of=start] {Convolutional Autoencoder Training};
\node (latent) [rectbox, fill=orange!30, below of=ae]  {Encoding Dataset Into Latent Sequences};
\node (graph)  [rectbox, fill=green!30, below of=latent]{Converting Latent Sequences to Graphs};
\node (gcn)    [rectbox, fill=blue!20, below of=graph]{Defining and Implementing Physics Loss};
\node (phys)   [rectbox, fill=orange!30, below of=gcn]{Building the GCN-LSTM Architecture};
\node (train)  [rectbox, fill=green!30, below of=phys] {Training and Evaluating the Model};
\node (decode) [rectbox, fill=blue!20, below of=train] {Decoding Predicted Latents to Images};
\node (vis)    [rectbox, fill=orange!30, below of=decode]{Visualization of Predictions};
\node (stop)   [rectbox, fill=red!30, below of=vis]    {Predicting Long-Horizon Microstructures};

% arrows
\draw [arrow] (start) -- (ae);
\draw [arrow] (ae) -- (latent);
\draw [arrow] (latent) -- (graph);
\draw [arrow] (graph) -- (gcn);
\draw [arrow] (gcn) -- (phys);
\draw [arrow] (phys) -- (train);
\draw [arrow] (train) -- (decode);
\draw [arrow] (decode) -- (vis);
\draw [arrow] (vis) -- (stop);

\end{tikzpicture}
\end{center}

The first model examined is the 2D Convolutional Autoencoder with input dimensions 128$\times$128. The construction of the pipeline and associated methods is described in detail for this model, while the subsequent two models are discussed only in terms of their differences to avoid redundancy. The entire modeling pipeline is structured to capture and forecast the temporal evolution of microstructure images using a hybrid deep learning framework. The process begins with loading and preprocessing raw microstructure images from Excel files. These images are then used to train a 2D convolutional autoencoder, which learns to compress each high-dimensional frame into a compact latent representation and reconstruct it with minimal loss of information. Once trained, the encoder component is applied across the dataset to transform image sequences into their corresponding latent representations. These encoded sequences serve as input for a graph-based modeling stage, where each latent frame is converted into a spatial graph, treating each pixel as a node with its associated 128-dimensional feature vector and local connectivity. The GCN-LSTM-PI model is then trained to predict the latent representation at the next timestep, given a sequence of graph-structured inputs. After prediction, the latent output is passed through the pre-trained decoder to reconstruct the corresponding image. This decoding step is crucial for both visual and quantitative evaluation, enabling predicted outputs to be compared with ground-truth images using pixel-wise metrics, such as mean squared error (MSE) and structural similarity index (SSIM). Finally, the framework is extended to perform long-range forecasts using a one-shot strategy. The model predicts and decodes the next latent state, and it compares it against multiple ground truth frames to evaluate the quality of its extrapolation.

\subsection{Data Preparation}\label{subsec4}
\begin{figure}[H]
\centering
\begin{minipage}[t]{0.49\textwidth}
    \centering
    \includegraphics[width=\linewidth]{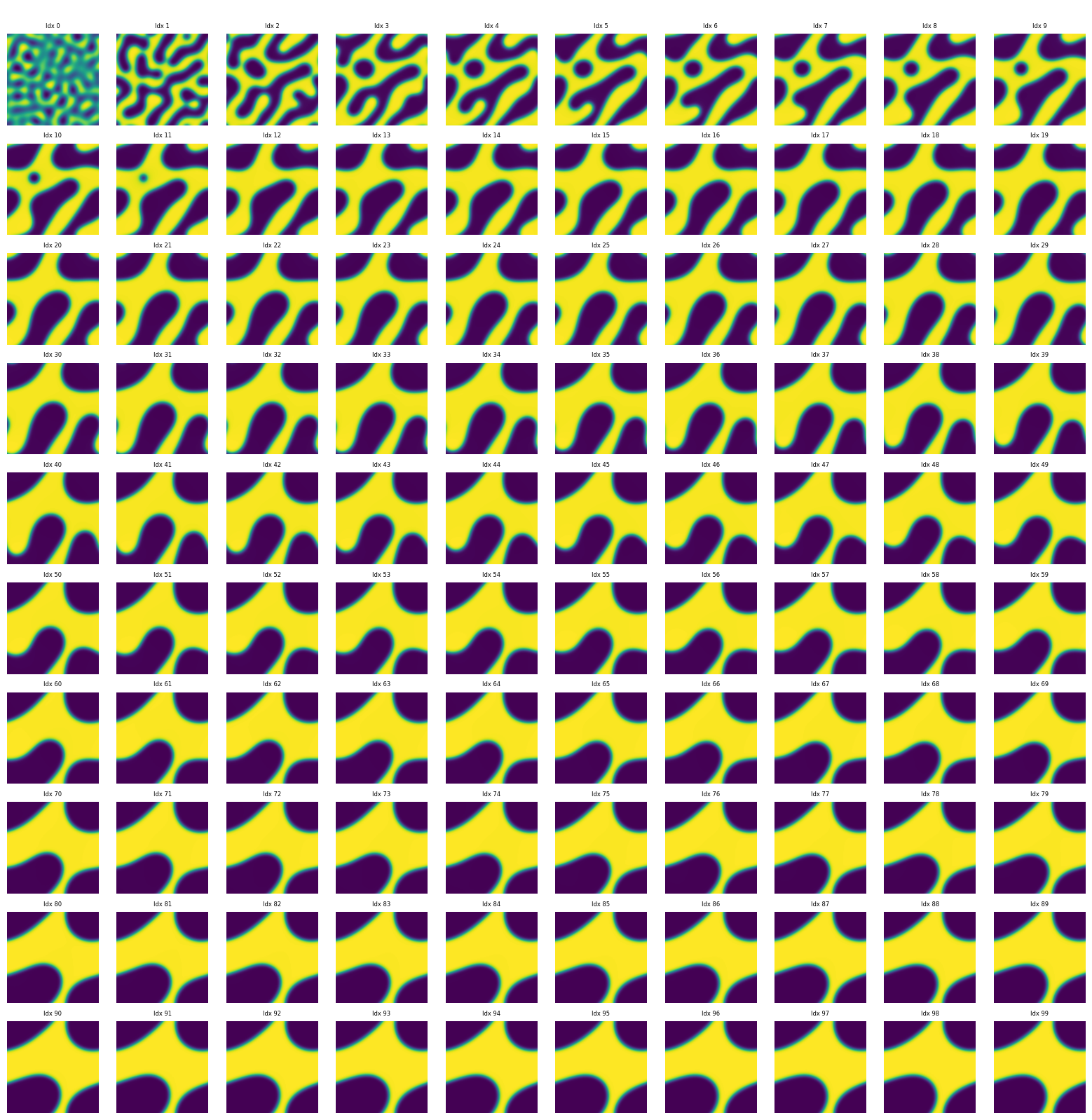}
    \subcaption{Sb composition: 0.5}
\end{minipage}
\hfill
\begin{minipage}[t]{0.49\textwidth}
    \centering
    \includegraphics[width=\linewidth]{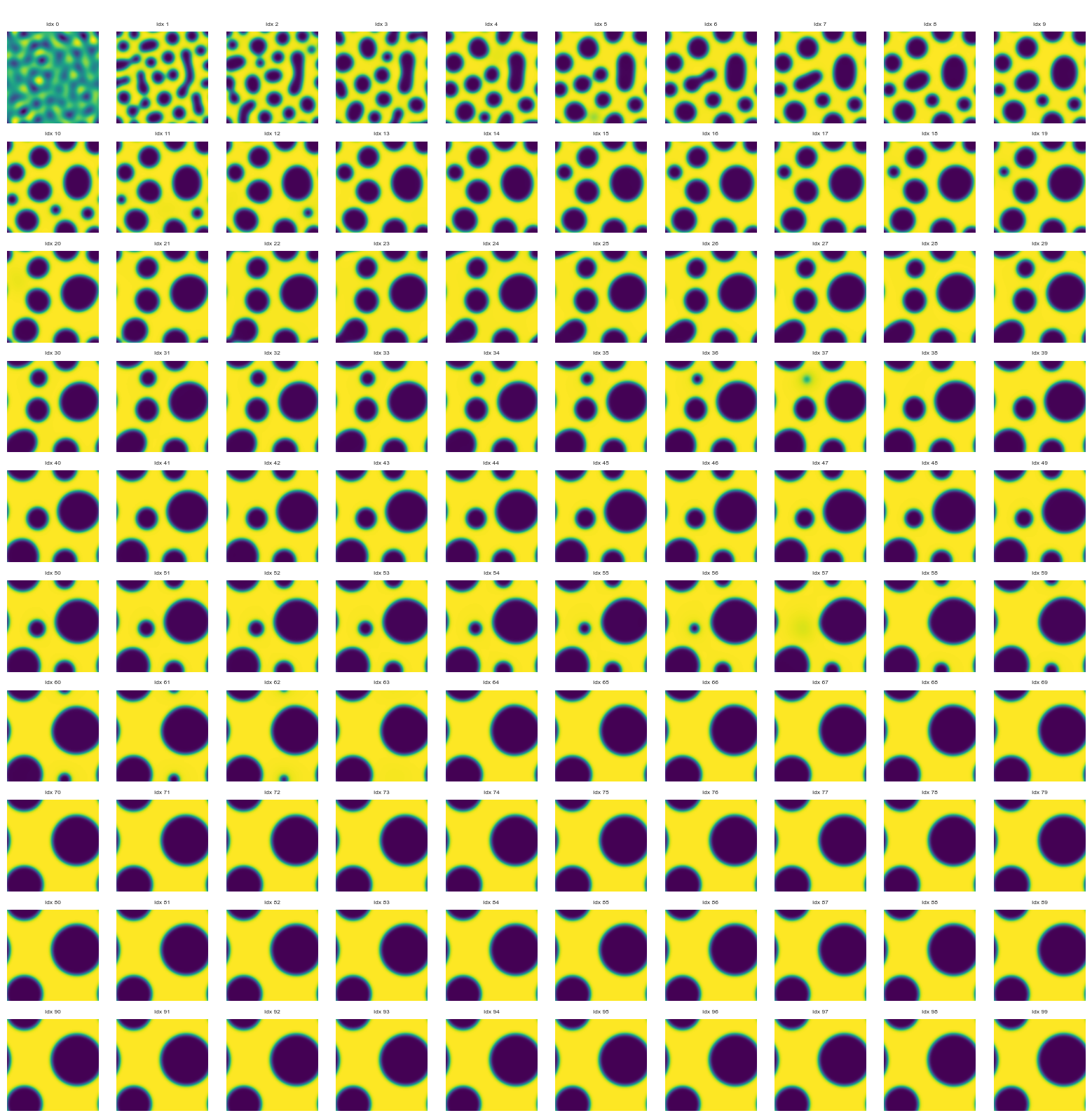}
    \subcaption{Sb composition: 0.6}
\end{minipage}
\caption{Microstructure datasets of Bi--Sb alloy were generated using the Cahn--Hilliard phase-field method at \(T=350\,\mathrm{K}\) with initial Sb compositions of 0.5 and 0.6. Each dataset contains \(128 \times 128\) images over 200{,}000 timesteps, with the first 100{,}000 used for training and the remaining 100{,}000 for forecasting evaluation. Forecasted microstructures are compared against unseen simulation data to assess long-term predictive capability. The joint dataset size is a total of 200 images.}
\label{fig:combined}
\end{figure}

To model the temporal evolution of microstructures, we first generated two datasets of simulation output via the Cahn-Hilliard phase-field method. The first dataset was generated at a temperature of 350 K and an initial Sb composition of 0.5. In contrast, the second dataset was generated at the same temperature with an initial Sb composition of 0.6. Each microstructure snapshot represents the molar fraction of antimony ($X_{\text{Sb}}$) in a Bi--Sb alloy. The images were saved as a $128 \times 128$ matrix in \texttt{.xlsx} format.

Each of the two datasets was simulated for a total of 200,000 timesteps. The first 100,000 timestep images were used for training, and the second 100,000 timestep images were used as unseen data to test the model's long-horizon forecasting capability. To capture the patterns of evolving microstructures, the first 100,000 timesteps of each dataset were used to train the model. Once training was complete, the model was assigned to forecast future images from timesteps 100,000 to 200,000. To evaluate the model's performance in forecasting, the forecasted images are compared against unseen datasets to test its ability to generalize and predict long-term microstructure dynamics. 

A custom data loader sequentially imported the image files from a designated folder path, converted them into NumPy arrays, and stacked them into a 4D array of shape $(N, 128, 128, 1)$, where $N$ represents the number of time steps. This array served as input to the autoencoder.

\begin{figure}[H]
\centering
\begin{minipage}[t]{0.49\textwidth}
    \centering
    \includegraphics[width=\linewidth]{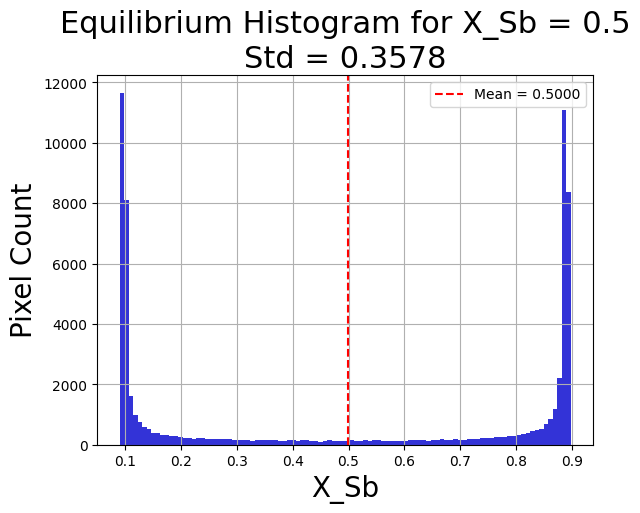}
    \subcaption{$\bar{X}_{\text{Sb}} = 0.5$}
\end{minipage}
\hfill
\begin{minipage}[t]{0.49\textwidth}
    \centering
    \includegraphics[width=\linewidth]{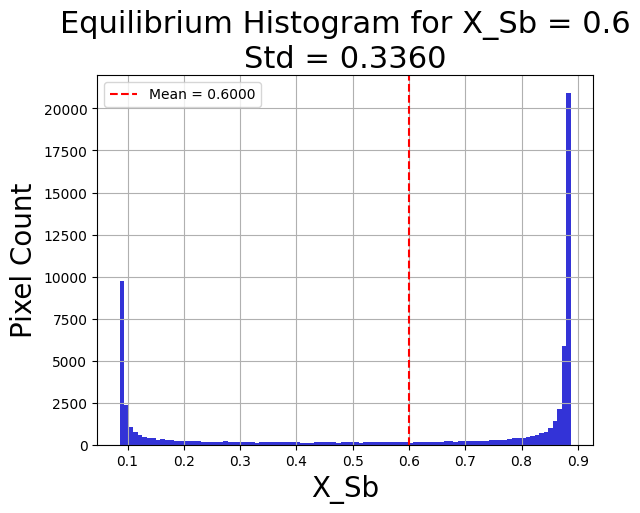}
    \subcaption{$\bar{X}_{\text{Sb}} = 0.6$}
\end{minipage}
\caption{Pixel-wise histograms of Sb composition at equilibrium for 0.5 and 0.6. Each distribution highlights variation, mean, and standard deviation in the simulated microstructures.}
\label{fig:combined}
\end{figure}

In the two histograms demonstrated above, the blue bars are the distribution of the composition ($X_{\text{Sb}}$) at the final time step 100,000 for the initial composition of antimony at 0.5 and 0.6, respectively. The dashed red lines represent the average composition. In the first plot ($\bar{X}_{\text{Sb}} = 0.5$), the histogram displays two symmetrical peaks centered at 0.1 and 0.9, indicating that the system has separated into two distinct phases with equal volume fractions. In contrast, the second plot ($\bar{X}_{\text{Sb}} = 0.6$) shows an asymmetric histogram, with the peak at approximately 0.9 being significantly taller than the peak at 0.1. Despite the system separating into the same two equilibrium phases, the Sb-rich domains occupy a larger volume fraction of the image due to their higher initial concentration. 

The next step in data preprocessing for our pipeline is to prepare the data for temporal modeling by defining a function that generates overlapping subsequences of a specific length $L$. The input is a tensor of shape $(N, 1, H, W)$, where $N$ is the number of time frames, $H$ and $W$ are the height and width of the image, respectively, as initially set in the phase field simulation. The function then returns a list of sequences, each with the shape of $(L, 1, H, W)$. These sequences serve as input for the model, allowing it to learn the microstructural evolution over time. The sequence lengths chosen in the pipeline were 3, 5, 7, and 10 to compare their effect on model training and generalizability in learning the dynamics of microstructure evolution.

To avoid data leakage, the sequences were first generated from the raw image datasets and subsequently randomly split into training and validation partitions of (80/20). This ensures that no overlapping frames are shared across training and validation. For each sequence length $L \in \{3, 5, 7, 10\}$, the sequences were generated separately from the $\bar{X}_{\text{Sb}} = 0.5$ and $\bar{X}_{\text{Sb}} = 0.6$ datasets. 
The randomly split training and validation sets were labeled with their respective composition and then merged into a joint dataset for each sequence length. To load the sequences along with their composition labels, a custom dataset class was created. Each sequence contains two channels, where the first channel encodes 
the local molar fraction of Sb at each pixel, and the second channel comprises a uniform composition map as a scalar that represents the global $\bar{X}_{\text{Sb}}$ value throughout the image. Consequently, a joint dataloader was built to combine the previously split sequences and construct PyTorch dataloader objects for the training and validation sets. For each sequence length $L$, the training and validation loaders were built using joint datasets of compositions 0.5 and 0.6. A configurable batch size was used, with a batch size of 8 applied in the pipeline. The loaders feed the model with batched sequences during training and evaluation, enabling scalable learning across various temporal contexts. 

Before moving on to the autoencoder architecture and training, a visual check is helpful to ensure the sequences are correctly constructed, preventing any mix-up between the two compositions during model training.  The visualization below, for a given sequence length, shows the training sequence $\bar{X}_{\text{Sb}} = 0.5$  in the upper row and $\bar{X}_{\text{Sb}} = 0.6$ in the lower row for $L = 5$. Each column displays a timestep to facilitate a visual comparison of how microstructures evolve for two compositions, and is repeated for all four sequence lengths. For each sequence length $L$, the visualizations are random due to a previous shuffle during the training and validation split. In the microstructure images, darker areas represent a lower concentration of Sb in that area.  

\begin{figure}[H]
\centering
\includegraphics[width=0.8\linewidth]{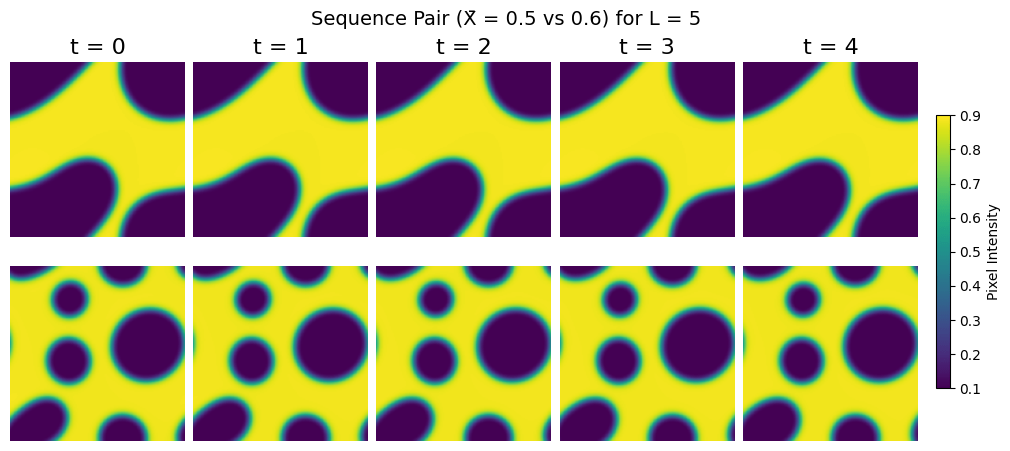}
\caption{Visualization of sequences for Bi--Sb microstructures with \(\bar{X}_{\mathrm{Sb}}=0.5\) (top row) and \(\bar{X}_{\mathrm{Sb}}=0.6\) (bottom row) at sequence length \(L=5\). Each column corresponds to a timestep, illustrating the temporal evolution of microstructures for the two compositions. Darker regions indicate lower Sb concentration.
}
\label{fig:sequence-pair}
\end{figure}

\subsection{2D Convolutional Autoencoder Training}\label{subsec5}

To facilitate dimensionality reduction in our dataset and support efficient spatio-temporal modeling, a 2D Convolutional Autoencoder architecture is implemented using PyTorch \cite{cheng2018deep}. The original microstructure images had high pixel dimensions of $128\times128$  pixels, and compressing them would drastically reduce the computational load. The autoencoder architecture utilizes a skip connection, along with explicit mean conservation enforcement on the microstructure channel, to ensure that mass is conserved throughout the training process. The skip connection here serves an essential purpose, helping the decoder recover fine-grained spatial details lost during downsampling \cite{ronneberger2015u}. The input has two channels, similar to those explained in the data preparation section.  

The encoder part of the model consists of two convolutional blocks. The spatial resolution $128\times128$ in the first block is reduced to $64\times64$, achieved using a convolutional layer with a stride 2, followed by batch normalization and ReLU activation \cite{ioffe2015batch}. In the second block, the spatial dimension is maintained and is not further downsampled to lower dimensions. Previously, the entire pipeline was trained on reduced dimensions $32\times32$, and the GCN-LSTM-PI model was unable to generalize well to the evolution of microstructures. Despite preserving the spatial dimension in $64\times64$, the feature maps are projected into a higher-dimensional latent space with 256 channels. The skip connection is maintained from the output of the first block and is reused during the decoding process. 

The decoder in our architecture mirrors the encoder structure in reverse order. The latent tensor is first transformed via a convolutional transpose layer and then concatenated with the encoder skip connection. This representation is then upsampled to the original spatial resolution of $128\times128$ through a second transposed convolution. The use of additional convolutional layers helps further refine the reconstruction of microstructure images in a single-channel output format. Finally, a sigmoid activation ensures that the output values remain within the normalized range of [0, 1] \cite{rumelhart1986learning}. Once the architecture of the 2D convolutional autoencoder was established, the model went through a joint training on microstructure sequences with an average composition of $\bar{X}_{\text{Sb}} = 0.5$ and $\bar{X}_{\text{Sb}} = 0.6$ with two channels, which were previously defined. In this step, reconstruction of the microstructures was performed to verify the sanity of the 2D convolutional autoencoder.

To quantify the reconstruction capacity of the architecture, a combined loss of MSE, SSIM, and a mean-conservation term was used after training and validation. In the validation phase, reconstruction performance was evaluated separately for each composition to ensure proper generalization. The training results are presented in Table 1.  The training phase was completed over 100 epochs, with a batch size set to 8 and the Adam optimizer configured with a learning rate of \(1 \times 10^{-4}\) for each sequence length. The Mean Squared Error (MSE) is defined as

\begin{equation}
\text{MSE} = \frac{1}{N} \sum_{i=1}^{N} (x_i - \hat{x}_i)^2
\end{equation}
Where $\text{MSE}$ denotes the mean squared error, $N$ is the total number of pixels in the image, $x_i$ is the pixel intensity of the original input image at position $i$, and $\hat{x}_i$ is the corresponding pixel intensity in the reconstructed image. The MSE penalizes larger deviations more strongly by squaring the difference, encouraging the network to produce reconstructions that closely match the input \cite{hinton2006reducing}.

\begin{equation}
\text{SSIM}(x, \hat{x}) = \frac{(2\mu_x \mu_{\hat{x}} + C_1)(2\sigma_{x\hat{x}} + C_2)}{(\mu_x^2 + \mu_{\hat{x}}^2 + C_1)(\sigma_x^2 + \sigma_{\hat{x}}^2 + C_2)}
\end{equation}
SSIM computes the structural similarity index between the original image $x$ and the reconstructed image $\hat{x}$ \cite{wang2004image}. $\mu_x$ and $\mu_{\hat{x}}$ are the mean pixel intensities of $x$ and $\hat{x}$, respectively. $\sigma_x^2$ and $\sigma_{\hat{x}}^2$ represent the variances, and $\sigma_{x\hat{x}}$ is the covariance between $x$ and $\hat{x}$. The constants $C_1$ and $C_2$ are used to stabilize the division, typically defined as $C_1 = (K_1 L)^2$ and $C_2 = (K_2 L)^2$, where $L$ is the dynamic range of the pixel values and $ K_1$ and $ K_2$ are small constants. SSIM ranges from 0 to 1, with higher values indicating greater structural similarity.

\begin{table}[h]
\caption{Validation metrics for varying sequence lengths at 100th epoch for $\bar{X}_{\mathrm{Sb}} = 0.5$ and $\bar{X}_{\mathrm{Sb}} = 0.6$ for the 2D Convolutional Autoencoder model with input size $128 \times 128$}\label{tab:sequence_metrics}
\begin{tabular*}{\textwidth}{@{\extracolsep\fill}lcccccc}
\toprule%
& \multicolumn{3}{@{}c@{}}{\textbf{$\mathbf{\bar{X}_{\mathrm{Sb}}} = \mathbf{0.5}$}} & \multicolumn{3}{@{}c@{}}{\textbf{$\mathbf{\bar{X}_{\mathrm{Sb}}} = \mathbf{0.6}$}} \\
\cmidrule{2-4}\cmidrule{5-7}
\textbf{Seq. Length} & \textbf{Valid. MSE} & \textbf{Valid. SSIM} &  
& \textbf{Valid. MSE} & \textbf{Valid. SSIM} & \\
\midrule
3  & 0.000008 & 0.999471 & & 0.000005 & 0.999542 & \\
5  & 0.000014 & 0.999368 & & 0.000008 & 0.999341 & \\
7  & 0.000006 & 0.999317 & & 0.000004 & 0.999535 & \\
10 & 0.000005 & 0.999341 & & 0.000005 & 0.999530 & \\
\botrule
\end{tabular*}
\end{table}

\begin{figure}[H]
\centering
\includegraphics[width=1\linewidth]{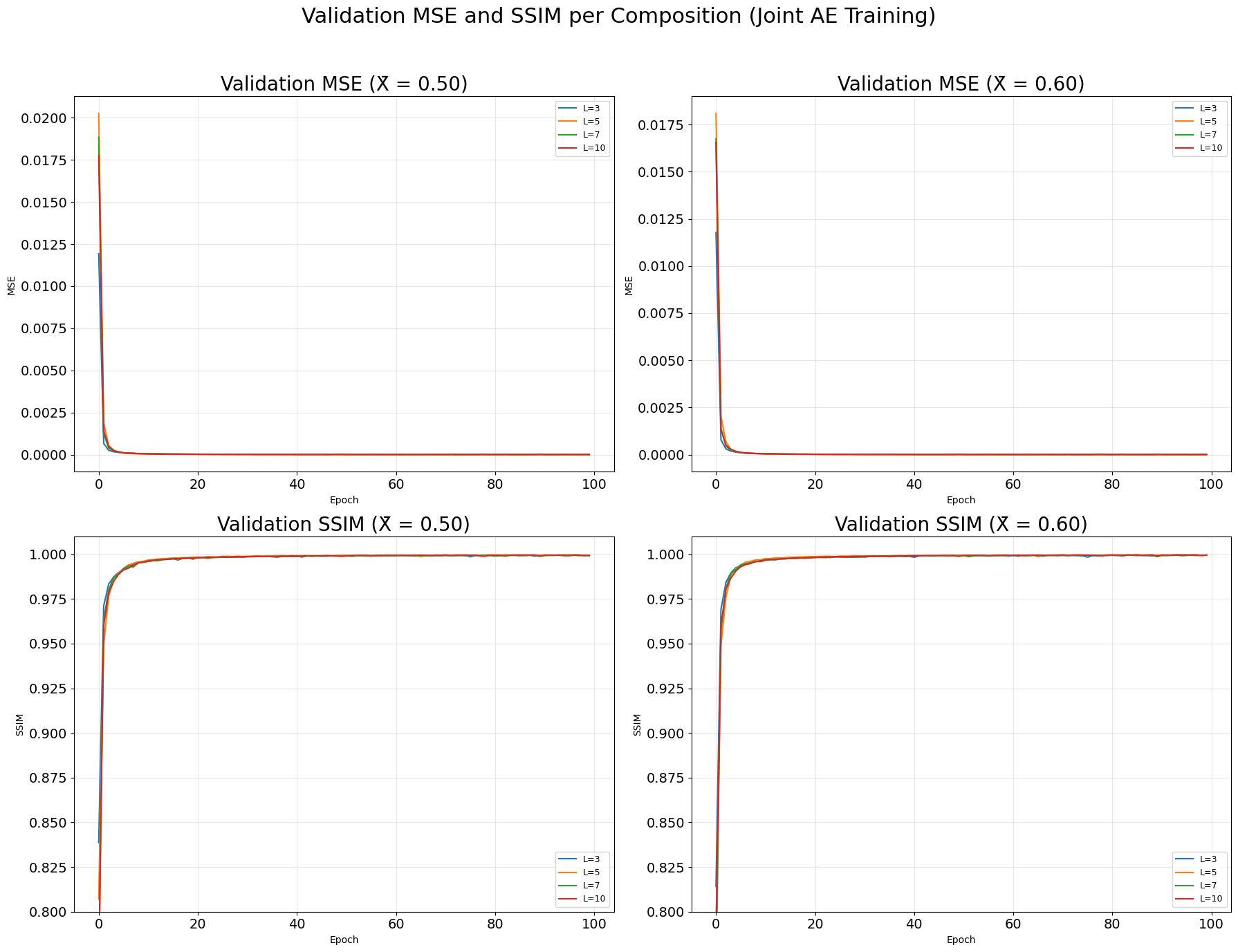}
\caption{Visualization of the validation performance for the 2D convolutional autoencoder across sequence lengths \(L \in \{3,5,7,10\}\) and input size $128 \times 128$ over 100 epochs. Validation MSE decreases and SSIM increases rapidly within the first 10 epochs, stabilizing thereafter. All sequence lengths and compositions (\(\bar{X}_{\mathrm{Sb}}=0.5\) and 0.6) converge to similarly low MSE and high SSIM values, indicating robust generalization. }
\label{fig:validation-metrics}
\end{figure}

The table and plots above show the performance of the 2D convolutional autoencoder in all four sequences \( L \in \{3, 5, 7, 10\} \). Within the first 10 epochs, the validation MSE decreases rapidly and the SSIM increases quickly, followed by stabilization of both metrics for the remaining 90 epochs. For both 0.5 and 0.6 compositions, all four sequence lengths converge to similarly low MSE and high SSIM values with minimal differences. This indicates that the autoencoder can generalize well for all sequence lengths and compositions.

\begin{figure}[H]
\centering
\includegraphics[width=0.7\textwidth]{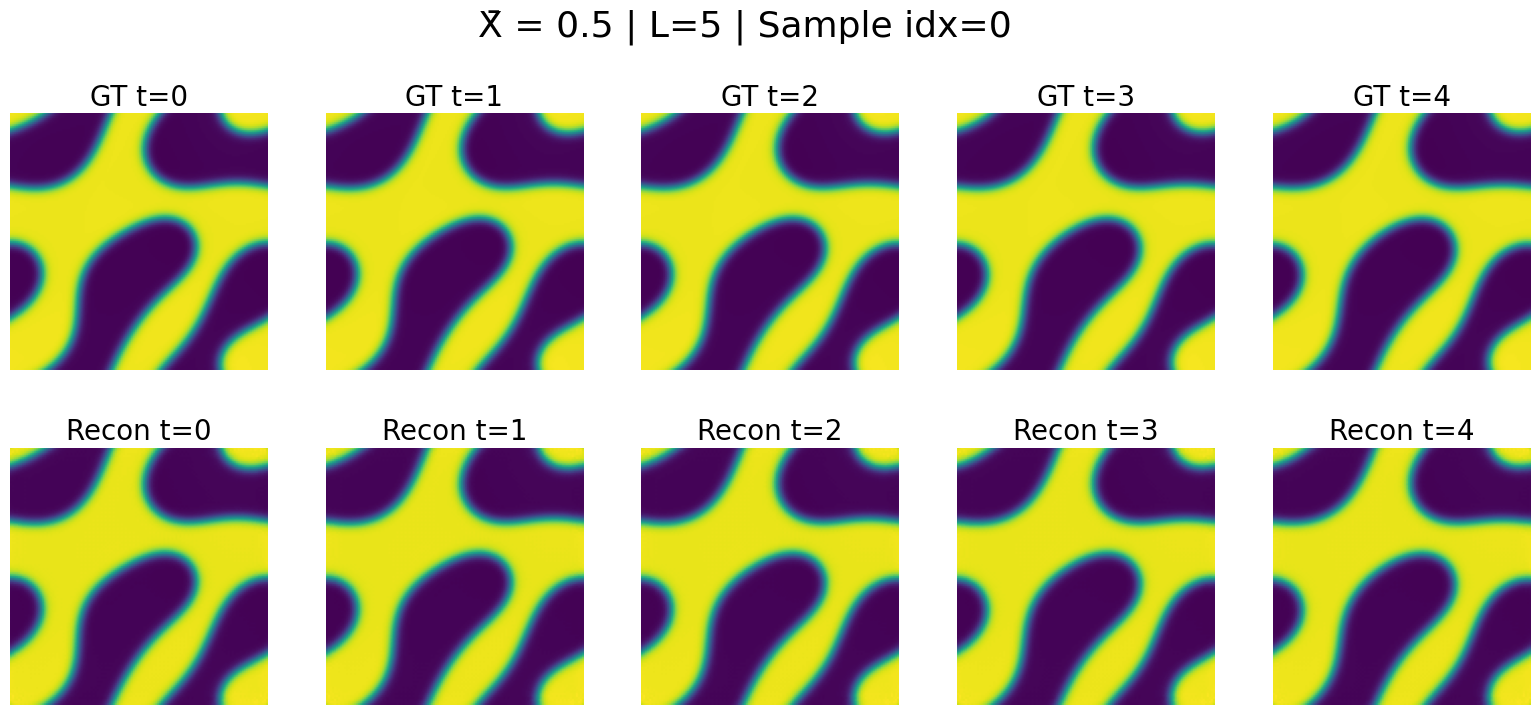}\\[0.5em]
\includegraphics[width=0.7\textwidth]{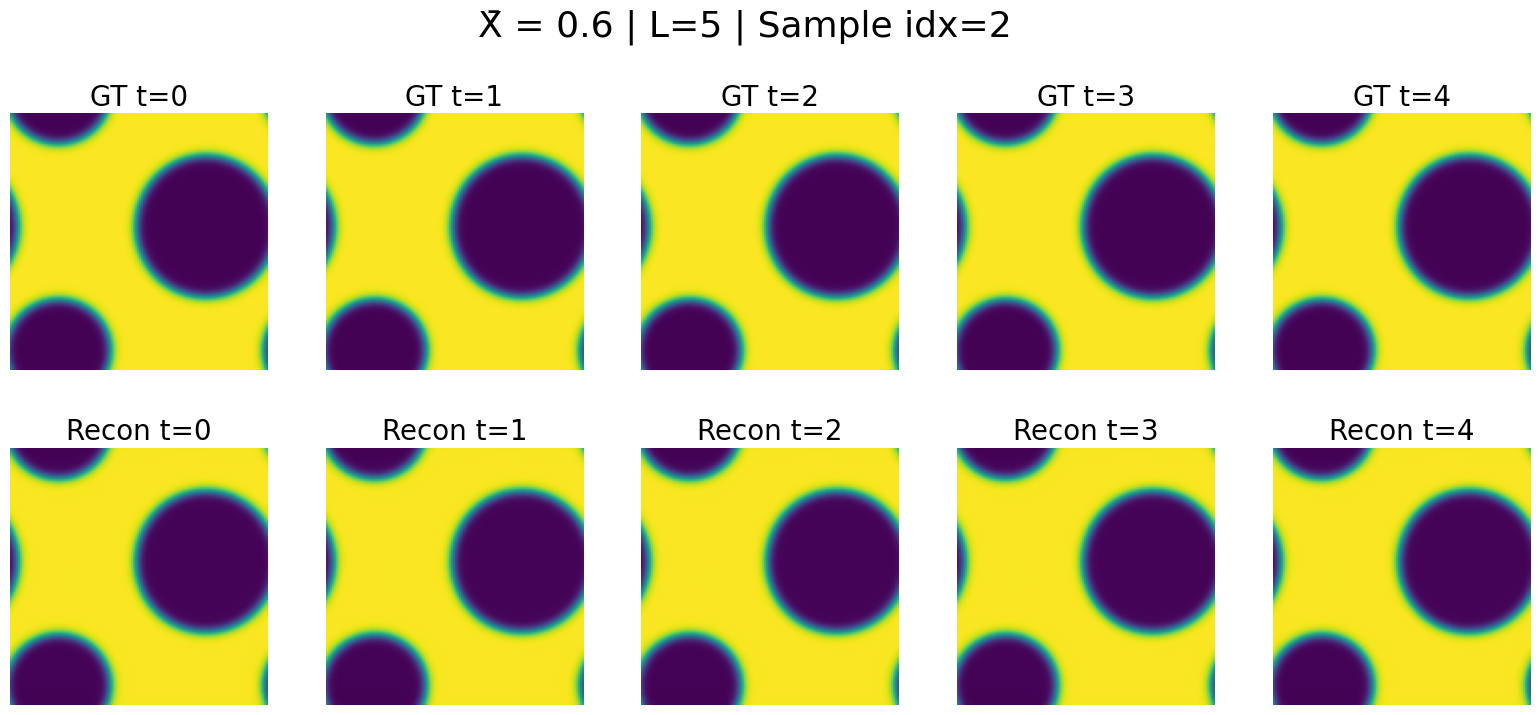}
\caption{Autoencoder reconstruction results for $L = 5$ at $\bar{X}_{\mathrm{Sb}} = 0.5$ (top) and $\bar{X}_{\mathrm{Sb}} = 0.6$ (bottom), showing both output sequences and encoded representations. GT stands for Ground Truth and Rec for Reconstructed.}
\label{fig:ae_recon_combined}
\end{figure}

The final step in autoencoder training is to visually verify the quality of the trained autoencoder reconstruction across different sequence lengths and compositions. In the above images, the sequence length of five was chosen for compositions 0.5 and 0.6 for visual verification. In this representation, validation samples are drawn, decoded, and then compared with the original images. It should be noted that this is not the final reconstruction used in the evaluation of the GCN-LSTM-PI model. It serves only as a sanity check to confirm that the autoencoder is functioning correctly and can accurately decode input sequences. As can be seen, the autoencoder performed very well in encoding and decoding the microstructure images while also conserving the composition throughout the training. The visualizations in Figure 7 correspond to the original and reconstructed plots for L = 5, along with three latent channels. 

\begin{figure}[H]
\centering
\includegraphics[width=0.7\textwidth]{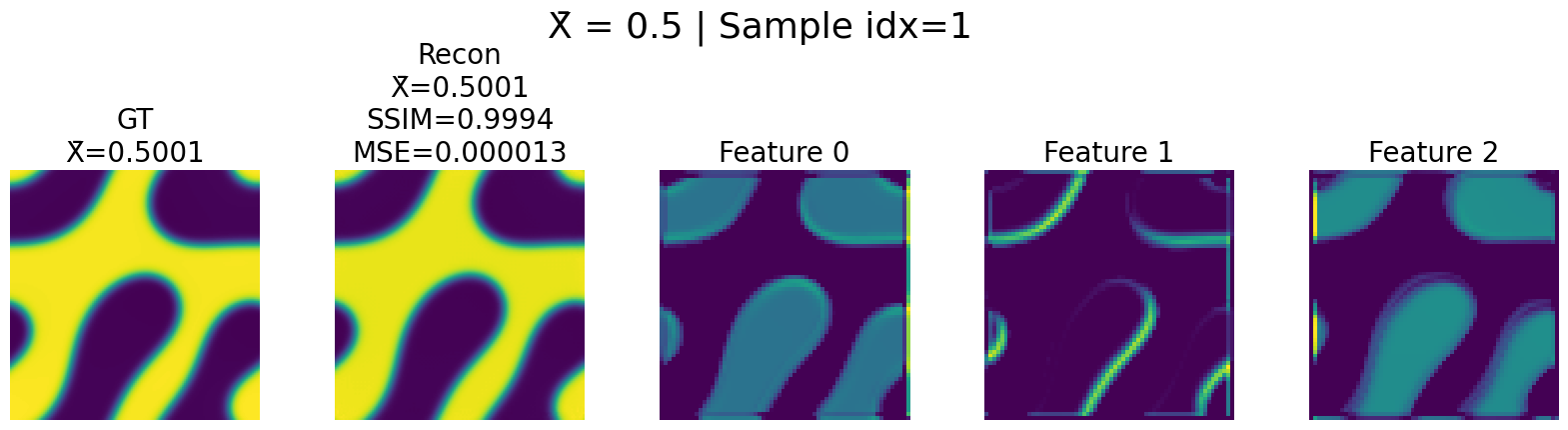}\\[0.5em]
\includegraphics[width=0.7\textwidth]{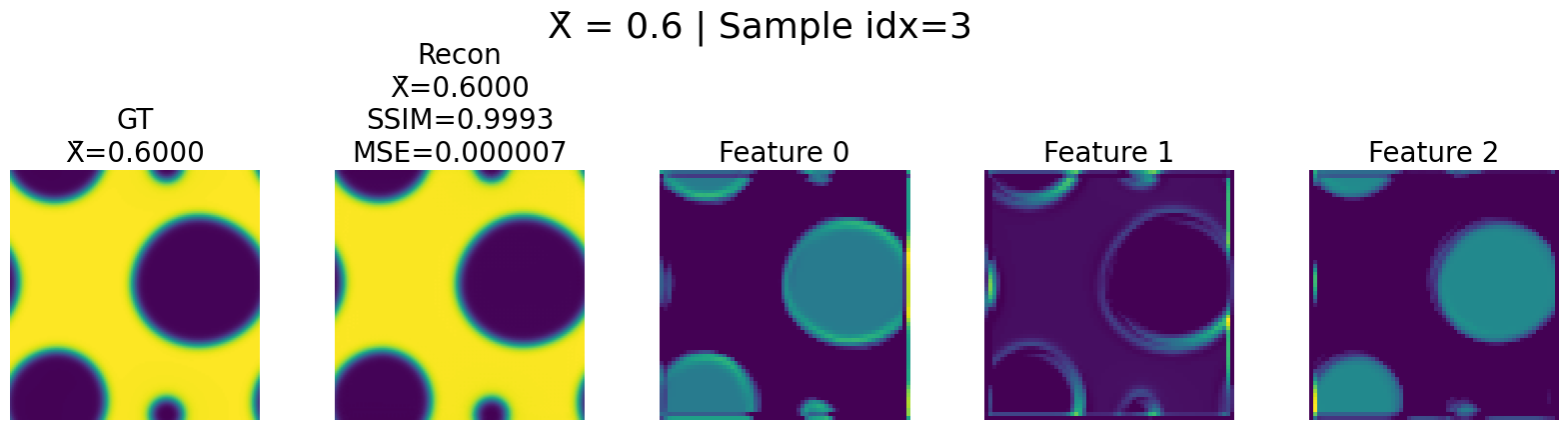}
\caption{2D autoencoder reconstruction results for input size $128 \times 128$ with sequence 
length $L=5$ for $\bar{X}_{\mathrm{Sb}} = 0.5$ (top) and $\bar{X}_{\mathrm{Sb}} = 0.6$ (bottom). 
For each composition, the first column shows the original microstructure (GT), the 
second column is the reconstructed output (Recon), followed by three representative latent 
filters. The reconstructed microstructures are evaluated using SSIM and MSE metrics.}
\label{fig:ae_recon_combined}
\end{figure}
\subsection{Encoding Dataset into Latent Sequences}\label{subsec6}

During training of the 2D convolutional autoencoder, the entire dataset is encoded, including the training and validation sets. The encoded dataset is converted to latent representations for each sequence length (3, 5, 7, 10). However, these latents are not stored during training. Therefore, in a separate step, the full training and validation sequences are re-encoded using the trained autoencoder. This allows the conversion of sequences to latents and saves them on disk to use them later in the downstream. Each input sequence has a shape of $(T, L, 2, 128, 128)$, where $T$ is the number of sequences, $L$ is the length of the sequence, and the two channels represent the microstructure and local composition. These were passed through the pre-trained autoencoder, producing latent tensors of shape $(T, L, 256, 64, 64)$. The output shapes confirm that each 128×128 frame was compressed into a 256-channel 64×64 representation, enabling more efficient downstream modeling. For instance, at a sequence length of 3, 396 sequences are obtained, and each is reduced to a latent size of (3, 256, 64, 64). The data were saved to disk for reuse in graph-based forecasting models. 

\subsection{Converting Latent Sequences to Graphs}\label{subsec7}

It is important to note that the raw input images are not used directly to build the graph. Instead, the graph structure and the node features are extracted from the latent representation learned by the encoder \cite{yang2024self}. This approach enables the encoded microstructural features to be embedded in a graph-based format that retains spatial locality and is well-suited for temporal modeling using Graph Convolutional Networks (GCNs) or hybrid GCN-LSTM architectures.

Transformation from a microstructure image to a graph begins with a two-channel input size of $128\times128$, where the first channel encodes the grayscale antimony concentration and the second channel contains a uniform composition map with a mean value of $\bar{X}_{\mathrm{Sb}} = 0.5$ or $\bar{X}_{\mathrm{Sb}} = 0.6$. As described previously, this input is passed through a 2D convolutional autoencoder, which compresses it into a latent feature tensor of shape $256\times64\times64$, where 256 denotes the number of feature channels, and $64\times64$ represents the downsampled spatial resolution.

To convert this latent tensor into a graph, each of the $64\times64$ spatial positions is treated as a node, resulting in a total of 4096 nodes. Each node is assigned a 257-dimensional feature vector by concatenating the 256 latent features at that location with the global scalar composition value. The edges are constructed by connecting each node to its four nearest neighbors, top, bottom, left, and right, forming a regular 4-connected  \cite{li2015gated}. This would result in localized message passing and spatial consistency in subsequent GCN-LSTM-PI processing. 
For visualization validation, a static one-time conversion of the latent images for each composition into graph structures is performed. In this setup, each node has a color that corresponds to the pixel intensity of the photos. The sole purpose of this visualization step is to confirm that the spatial patterns captured in the latent representation are transferable to the graph domain. 

Figure 8, compares raw microstructure images (left) with latent graphs colored by input intensity (right) for $\bar{X}_\mathrm{Sb}=0.5$ and $0.6$ at sequence length~5:

\begin{figure}[H]
\centering
\includegraphics[width=0.7\textwidth]{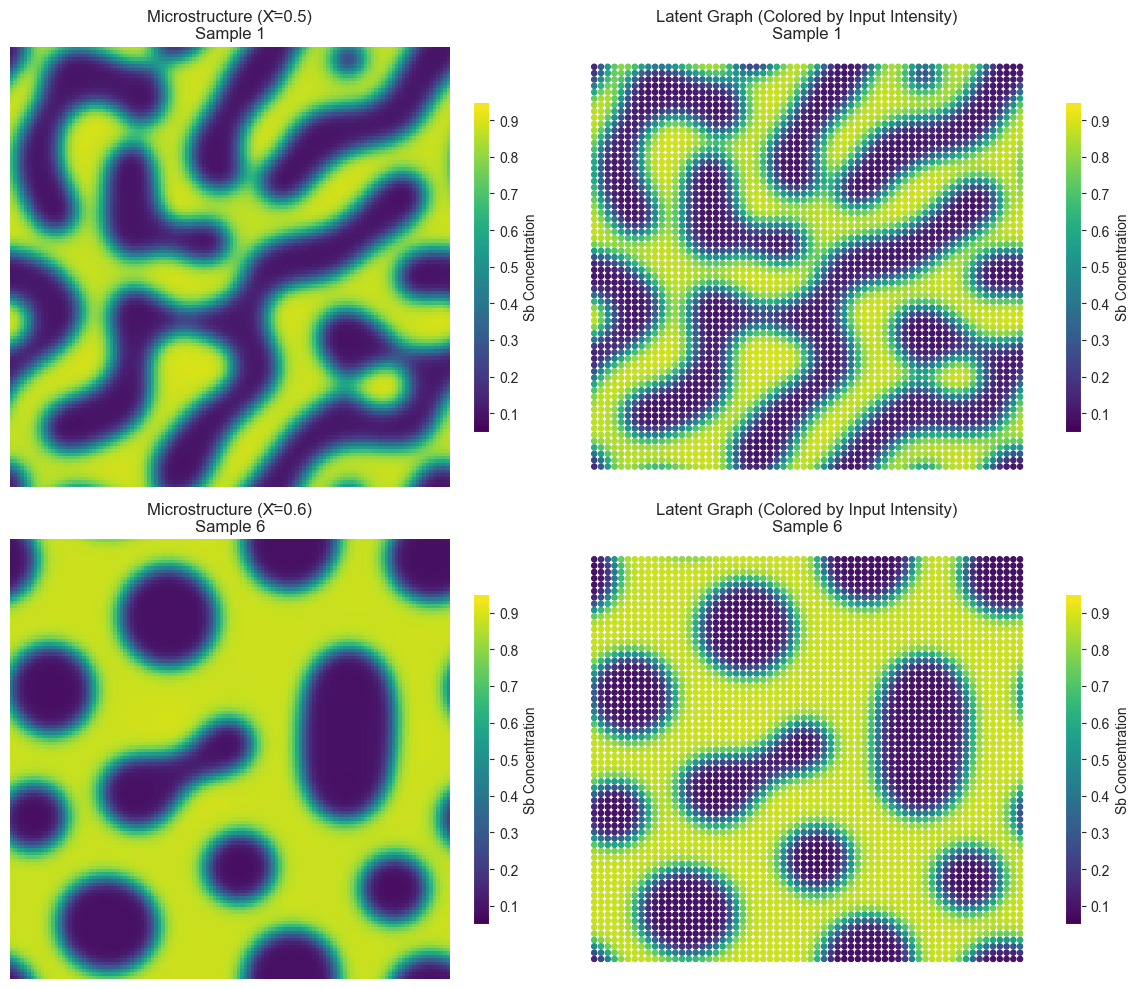}
\caption{Microstructure images are transformed into graphs through a latent encoding pipeline. 
A 2D convolutional autoencoder compresses each $128 \times 128$ two-channel input into 
a latent tensor of shape $256 \times 64 \times 64$, which is converted into a graph of 
$4096$ nodes with $257$-dimensional features ($256$ latent + $1$ composition).}
\label{fig4}
\end{figure}

Next, the latent sequences from the $(T-1)$ timesteps are converted into temporal graph sequences suitable for model training. The entire sequence of latent graphs is structured for temporal modeling. This step is facilitated through the \texttt{LazyGraphSequenceDataset}, which receives tensor sequences with the shape of \( (T-1)\times  C\times  H\times W \). The timesteps are then converted into graphs, where each node represents one spatial position and contains a feature vector comprising a latent encoding concatenated with the scalar composition value. The term lazy refers to the on-demand conversion of latent tensors to a graph sequence only when they are accessed. In practice, the latents of $T-1$ are converted into graphs, while the final frame $T$ is held out as the prediction target. This approach was used due to memory constraints and to eliminate the need to precompute and store all graph objects in memory \cite{fey2019fast}. In addition, target images also comprise two channels for microstructure and composition, transforming their shape from \(1 \times H \times W\) to \(2 \times H \times W\).  These targets are driven from the final frame, which ensures that the model learns to predict the next microstructure frame by having the preceding latent sequence. With this design, both the evolving spatial features and the global compositions are directly embedded in the inputs and targets of the model. 

To define how samples are batched during training, a \texttt{collate} function is defined that separates the list of \texttt{(graph\_sequence, target)} pairs into two parts. The first part retains graph sequences as lists to preserve temporal order, and the second part stacks target images into a single tensor \cite{wu2019graph}. This structure is essential for compatibility with PyTorch Geometric, which operates on individual graphs per timestep, allowing GCN-LSTM-PI to process graph-based time series data efficiently.   

After defining the \texttt{collate} function, a \texttt{Dataloaders} function is defined to allow large-scale and memory-efficient training \cite{hamilton2017inductive}. For each of the four sequence lengths, this function loads the corresponding latent graph sequences and raw microstructure images directly, computes the scalar composition from the final frame, and injects this composition as a second channel into the target tensor. The data is split into training and validation following the previous split performed in the pipeline, and it is used to create instances of \texttt{LazyGraphSequenceDataset}. Finally, everything is wrapped into PyTorch \texttt{DataLoaders} by utilizing the custom \texttt{collate}, which was described earlier. This modular structure enables the benchmarking of various sequence lengths and their performances, while maintaining a low memory overhead to run computations on a local computer.

Once the Dataloader building function is defined, it is invoked with the specified sequence lengths, pre-processed dataloaders, and along the defined encoded edge indices. Each entry would include a validation and a training loader that form the final pipeline for the GCN-LSTM-PI model. The purpose of loaders is to ensure that the encoded composition-aware temporal graphs are delivered efficiently into the model for both training and validation.  

\subsection{Defining and Implementing Physics Loss}\label{subsec8}

Implementation of the physics-informed loss ensures that the predicted microstructures are consistent with the underlying physics and is done by replication of the equations implemented in the phase-field simulation discussed earlier. The first step is to compute the molar volume of each sample based on the average antimony composition $X_{\text{Sb}}$.
Therefore, a function is defined that takes a 3D or 4D tensor representing batch images and computes the mean composition of each sample. This function uses a weighted average of the elemental molar volumes of Sb and Bi, with constants $V_\text{Sb} = 21.31 \times 10^{-6}$~m$^3$/mol and $V_\text{Bi} = 18.19 \times 10^{-6}$~m$^3$/mol to compute the molar volume. This computation yields a 1D tensor of molar volumes per batch, which is used in subsequent physics-based calculations. 

To evaluate whether the predictions are consistent with the Cahn-Hilliard dynamics, a batch-wise loss function is defined. This function enables the model to simulate the phase-field evolution of microstructures in our Fourier space dataset using semi-implicit time integration \cite{eyre1998unconditionally}. It takes the predicted concentration field $X_t$ along with the original images of the next timestep $X_{t+1}$, using the physical parameters that were defined, such as temperature, which is $T = 350$K, grid size $\Delta x = 1 \times 10^{-8}$~m, mobility $M = 1 \times 10^{-26}$~m$^5$/J/s, gradient energy coefficient $\kappa = 10^{-7}$J/m, and the universal gas constant $R = 8.3144$~J/mol/K. These parameters are used as constants in the model and can be adjusted on the basis of different scenarios. 

The next step in this process is to calculate the free energy derivatives $\partial G / \partial X$ based on the polynomial expression of $G_\text{Sb}(T)$ and $G_\text{Bi}(T)$, in addition to the parameters of the regular solution model $L_0$ and $L_1$. These derivatives are used to help obtain a driving force field $\partial G / \partial X$ that is normalized by the molar volume (divided by V)  and transformed into a Fourier space \cite{zhu1999coarsening}. Furthermore, spatial derivatives are represented by pre-computing squared and bi-Laplacian operators $\mathbf{g}^2$ and $\mathbf{g}^4$ using FFT wavevectors. The predicted concentration field at the next timestep is computed via a semi-implicit Fourier-space update:

\begin{equation}
\hat{X}_{t+1} = \frac{\hat{X}_t - M \Delta t \, \mathbf{g}^2 \hat{f}}{1 + \kappa M \Delta t \, \mathbf{g}^4}
\end{equation}

In equation 11, $\hat{X}_t$ and $\hat{f}$ are the Fourier transforms of the concentration and the free energy derivative fields, respectively. The resulting $\hat{X}_{t+1}$ is inversely transformed to obtain the predicted spatial domain field $X_{t+1}^{\text{pred}}$. The physics-informed loss determines how well our predicted microstructures align with the Cahn-Hilliard equation. It is computed as the normalized mean squared error between the predicted and true $X_{t+1}$ fields:

\begin{equation}
\mathcal{L}_{\text{physics}} = \frac{\|X_{t+1}^{\text{pred}} - X_{t+1}^{\text{true}}\|^2}{\|X_{t+1}^{\text{true}}\|^2}
\end{equation}

A crucial factor based on physical laws is the conservation of mass in the \cite{shen2013mass}. To facilitate this, a global mass conservation is applied to the predicted microstructures, and a loss function for the mass conservation is also applied within the training loop. Conservation loss is defined as the squared difference in mean values between the predicted and original concentration fields. In the equation below \( \bar{x}_i \), and \( \bar{\hat{x}}_i \) are the mean compositions of the original and predicted fields for the sample \( i \). This loss penalizes deviations from the expected mean to ensure consistency with physical laws. 

\begin{equation}
\mathcal{L}_{\text{conservation}} = \frac{1}{B} \sum_{i=1}^B \left( \bar{x}_i - \bar{\hat{x}}_i \right)^2
\end{equation}

\subsection{Building the GCN-LSTM Architecture}\label{subsec9}

This architecture is based on a hybrid structure that incorporates a \textbf{Graph Convolutional Network} with \textbf{Long Short-Term Memory} and is further \textbf{Physics-Informed}, which is referred to as \textbf{GCN-LSTM-PI} for short. A sequence of graphs, one per time step, is fed into the model as input, and the model outputs the predicted latent representation of the next time step, which is decoded into a microstructure image using the pre-trained decoder. The input to this model consists of a batch of sequences, and each graph contains 4,096 nodes, which is derived from multiplying the compressed height and width dimensions of $64\times64$. For each node, 257 features are assigned, where 256 channels come from encoder features, and one channel represents a scalar composition value. The edges of each graph follow a fixed 4-neighbor spatial connectivity pattern and are consistent across all graphs in the sequence. 
  
Per each timestep \( t \), the model processes the node features for each graph independently using a stack of three \texttt{GCNConv} layers. These layers aggregate local information by passing messages to capture spatial dependencies \cite{wu2020comprehensive}. The message passing in our GCN implementation is bidirectional, and the features are exchanged between the connected nodes at each layer. These messages are high-dimensional vectors derived from the latent autoencoder space \cite{gilmer2017neural}, and capture the local composition, texture, and physical state of the material at each location. The GCN performs the process of passing messages over the graphs, where each node aggregates information from its neighbors and updates its feature vector. To update the features at a given node, a weighted combination of its features and immediate neighbors is computed and normalized. This allows the network to capture the spatial dependencies and local interactions in microstructures that are essential for modeling spinodal decomposition. In the GCN architecture, each of the three layers performs the transmission of messages using a normalized adjacency matrix $\hat{A} = D^{-1/2}(A + I)D^{-1/2}$, where $D$ is the node degree matrix, and $I$ adds self-loops, and PyTorch Geometric internally implements this \cite{kipf2016semi}. Each layer performs batch normalization (\texttt{BatchNorm1d}) \cite{he2016deep} , and \texttt{ReLU activation} to stabilize and regularize the learning process. After the \texttt{GCNConv} stacks are applied to each timestep, a sequence of node embeddings is obtained with shape \( (B, L, N, F) \), where \( B \) is the batch size, \( L \) is the sequence length, \( N \) is the number of nodes, and \( F \) is the GCN hidden dimension (default 512) that is temporally ordered.

The hybrid LSTM system in our model processes each node independently over time steps, producing a sequence of outputs for each time step \cite{seo2018structured}. The input to the LSTM part is a reshaped sequence of \((B \cdot N, L, F)\), allowing the LSTM to capture the dynamics of node evolution sequentially. The architecture uses the output at the final step and yields a tensor of shape \( (B \cdot N, H) \), where \( H \) is the hidden size of LSTM (default 512). For regularization, \texttt{LayerNorm} \cite{ba2016layer} and \texttt{Dropout} \cite{srivastava2014dropout} are applied to the output, followed by a linear projection that reduces the dimensionality to 256 channels, thus forming the predicted latent representation. Ultimately, to obtain an output that matches the spatial resolution of the original encoded latent tensor, it is reshaped to $(B, C, H, W)$, where $C = 256$ and $H = W = 64$. This output is passed to the decoder for reconstruction of the predicted microstructure image. The GCN-LSTM-PI model is effective in capturing the spatial and temporal patterns in evolving microstructures, making it suitable for learning their dynamics. 

Another implementation of this 2D model was run with the $256\times256$ inputs instead of $128\times128$ to capture finer spatial details and a larger number of grains within each microstructure. Although computationally more demanding, the larger input provides richer morphological information for more accurate learning and alloy design. Since this model primarily utilizes the same steps as the previous model, only the major differences in the two pipelines will be discussed here. The modification of the pipeline was implemented to facilitate larger input images of size $256\times256$ compared to the earlier $128\times128$.  The first difference comes from using a different dataset generated by the phase-field simulation. In contrast to the previous model, which was trained in timesteps from 0 to 100,000, the $256\times256$ version was trained over timesteps between 0 and 20,000. The dynamics of microstructure evolution in this time window are fundamental in the design of alloys because they encompass the early stages of spinodal decomposition.  In this stage, numerous grains and interfaces exist, along with detailed information on the coarsening dynamics. Training the model in this window enables a richer and more diverse learning behavior. Therefore, it could potentially enhance the capacity of the model to generalize to the practical processing of alloys under conditions where the early-stage morphology often dictates the material's performance and characteristics. 

\begin{figure}[H]
\centering
\begin{minipage}{0.48\textwidth}
    \centering
    \includegraphics[width=\linewidth]{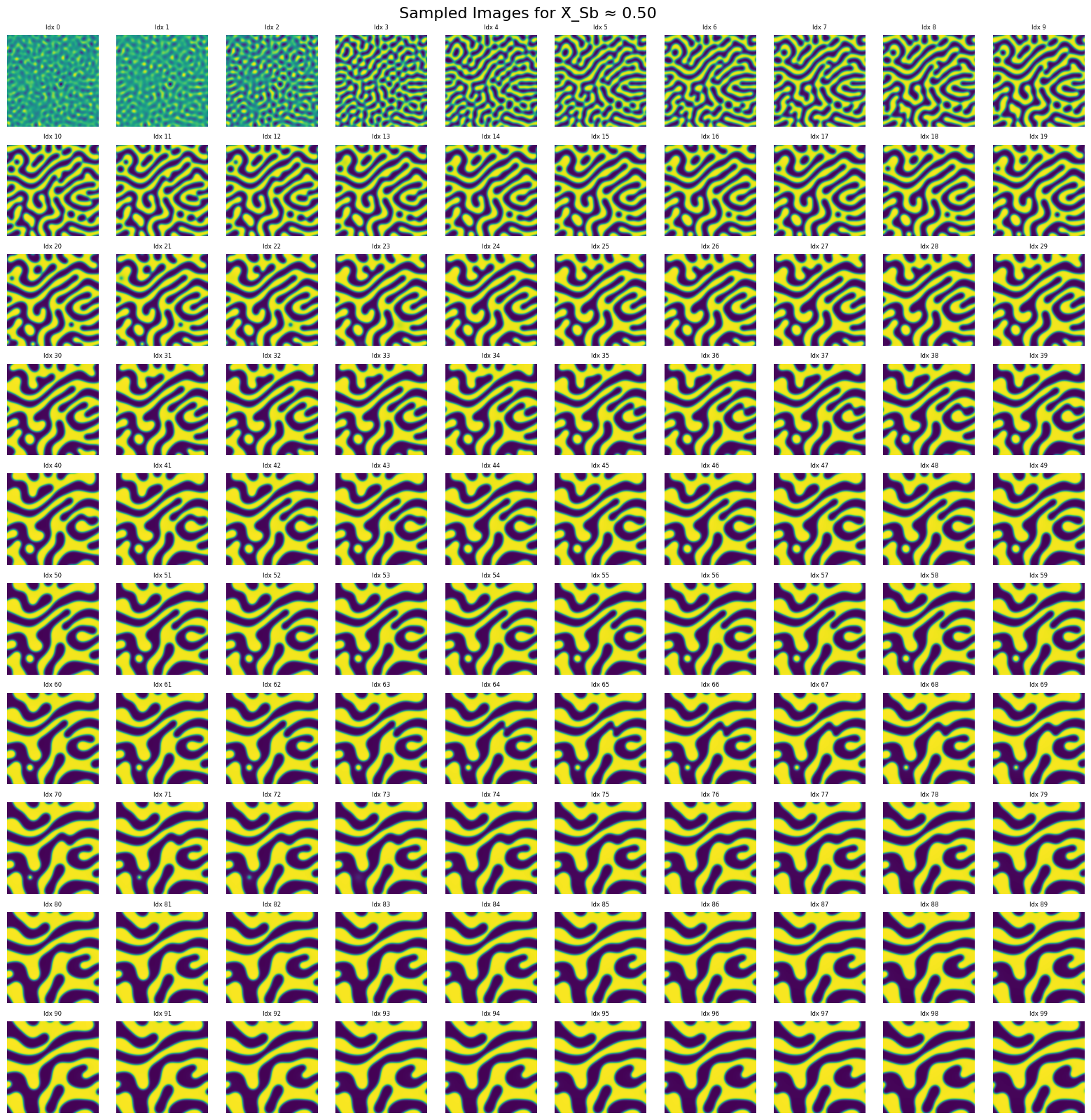}
\end{minipage}
\hfill
\begin{minipage}{0.48\textwidth}
    \centering
    \includegraphics[width=\linewidth]{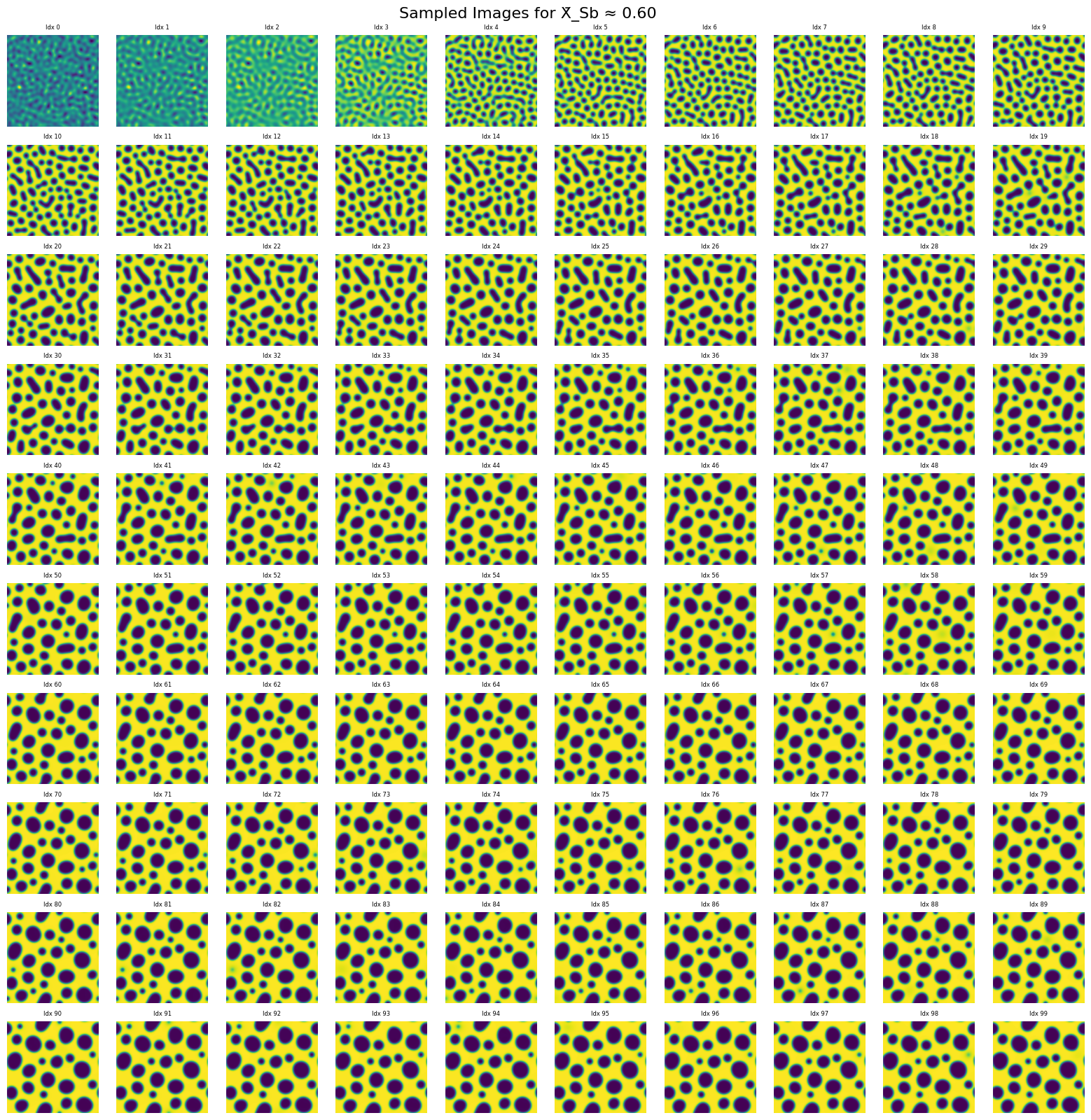}
\end{minipage}
\caption{Input microstructure images extracted from phase-field simulations at 350 K with an initial Sb composition of 0.5 (left) and 0.6 (right) and a total of 20,000 timesteps for training. The subsequent range from 20,000--40,000 timesteps is reserved as unseen data to evaluate forecasting accuracy. The joint dataset size is a total of 200 images with input size of $256\times256$.}
\label{fig: combined}
\end{figure}

For this model, the only sequence length used for both training and validation is $L = 3$. The same type of 2D Convolutional Autoencoder was used for this model, with some differences compared to the previous model to improve its capacity. The two architectures differ primarily in terms of depth, skip-connection design, and spatial resolution handling. The previous design downsamples from 128 to 64 resolutions, comprises two encoding and four decoding layers, and uses one skip connection. In contrast, the $256\times256$ model downsamples from 256 to 128 resolution,  uses five GCN layers, comprises three encoding and five decoding layers, and employs multiple skip connections. 
\begin{table}[h]
\caption{Validation metrics for varying sequence lengths at 100th epoch for $\bar{X}_{\mathrm{Sb}} = 0.5$ and $\bar{X}_{\mathrm{Sb}} = 0.6$ or the 2D Convolutional Autoencoder model with input size $256 \times 256$}\label{tab:sequence_metrics}
\begin{tabular*}{\textwidth}{@{\extracolsep\fill}lcccccc}
\toprule%
& \multicolumn{3}{@{}c@{}}{\textbf{$\mathbf{\bar{X}_{\mathrm{Sb}}} = \mathbf{0.5}$}} & \multicolumn{3}{@{}c@{}}{\textbf{$\mathbf{\bar{X}_{\mathrm{Sb}}} = \mathbf{0.6}$}} \\
\cmidrule{2-4}\cmidrule{5-7}
\textbf{Seq. Length} & \textbf{Valid. MSE} & \textbf{Valid. SSIM} &  
& \textbf{Valid. MSE} & \textbf{Valid. SSIM} & \\
\midrule
3  & 0.000005 & 0.998797 & & 0.000006 & 0.998943 & \\
 \\
\botrule
\end{tabular*}
\end{table}
\begin{figure}[H]
\centering
\includegraphics[width=0.9\textwidth]{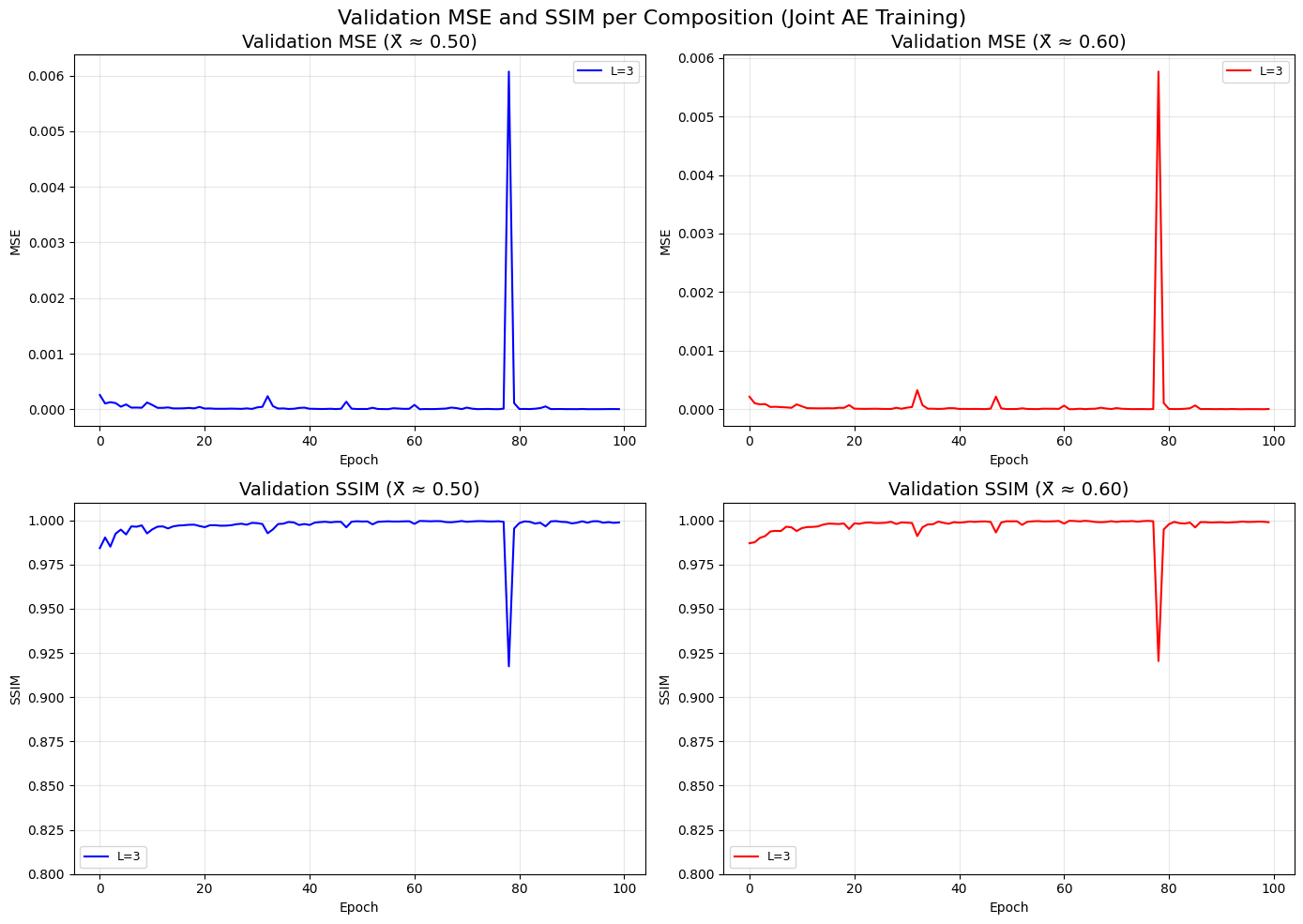}
\caption{Validation performance of the 2D convolutional autoencoder with sequence length $L=3$ and input size $256 \times 256$. 
The plots show the evolution of MSE (top) and SSIM (bottom) across 100 epochs for 
compositions $\bar{X}_{\mathrm{Sb}} = 0.5$ and $\bar{X}_{\mathrm{Sb}} = 0.6$.}
\label{fig4}
\end{figure}
Visualizations at sequence length $L=3$ for compositions 0.5 and 0.6 are shown in Figure 10 with their feature channels. Reconstruction quality (MSE, SSIM) improves compared to the $128 \times 128$ model.

\begin{figure}[H]
\centering
\includegraphics[width=0.8\textwidth]{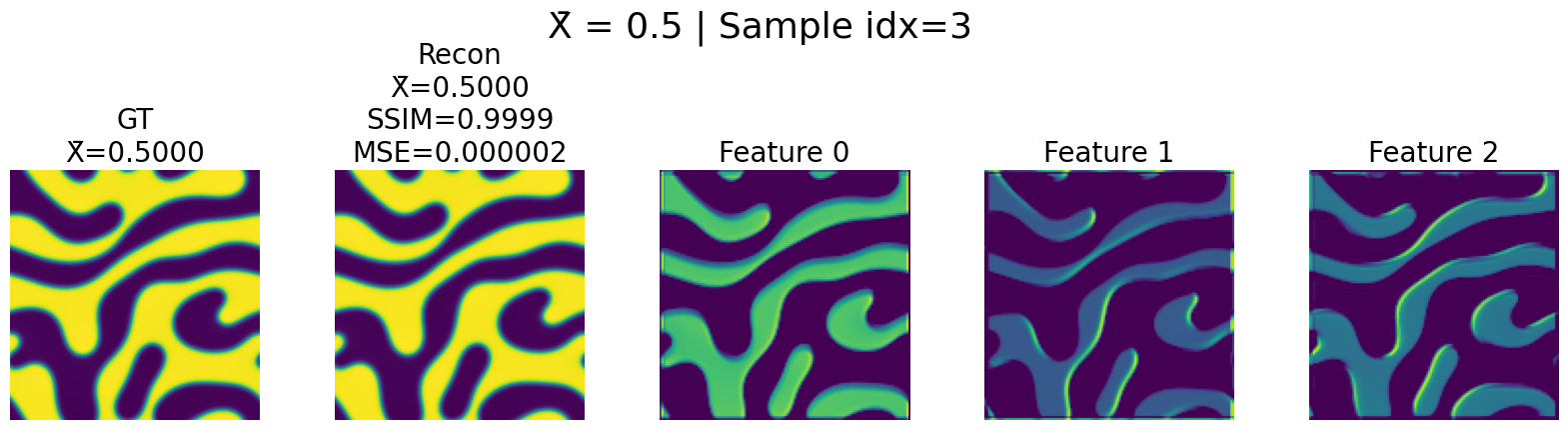}\\[0.5em]
\includegraphics[width=0.8\textwidth]{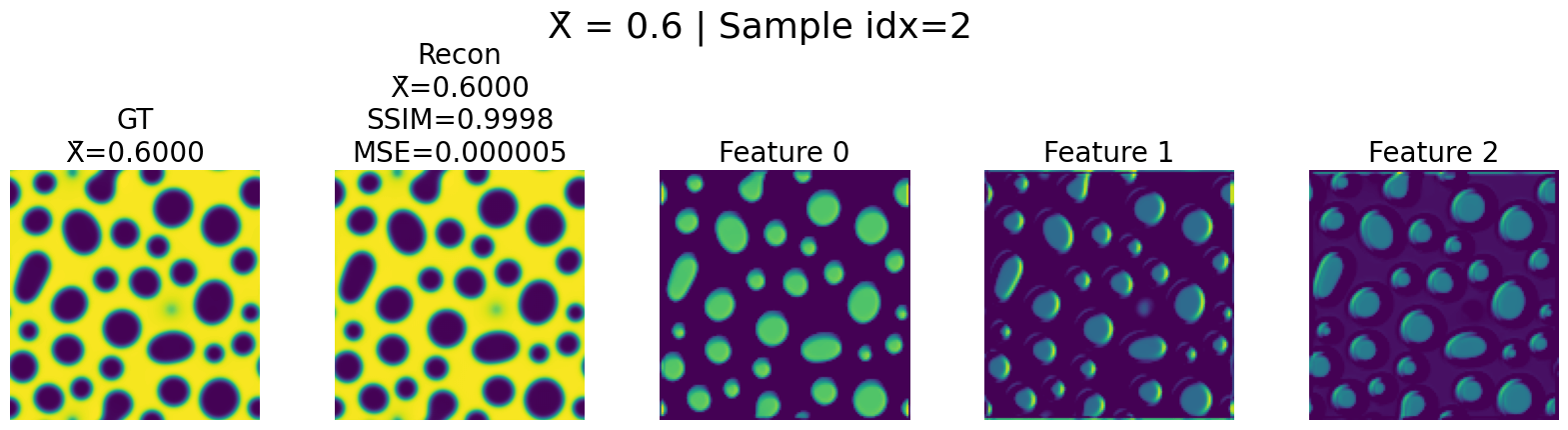}
\caption{2D autoencoder reconstruction results for input size $256 \times 256$ with sequence 
length $L=3$ for $\bar{X}_{\mathrm{Sb}} = 0.5$ (top) and $\bar{X}_{\mathrm{Sb}} = 0.6$ (bottom). 
For each composition, the first column shows the original microstructure (GT), the 
second column is the reconstructed output (Recon), followed by three representative latent 
filters.}
\label{fig:ae_recon_combined}
\end{figure}

In addition to pixel (2D) representations with resolutions of 128$\times$128 and 256$\times$256, a voxel (3D) representation with input size 128$\times$128$\times$128 was used to capture the complete spatial morphology of the microstructures. This approach enables the pipeline to learn physically consistent 3D evolution dynamics beyond what 2D projections can provide. Although the backbone of the pipeline remains the same as in 2D versions, significant modifications are essential to implement the 3D input.

For the 3D pipeline, volumetric data were loaded from \verb|.mat| files, since MATLAB conveniently preserves multidimensional voxel arrays without loss of structure. While the 2D setup encodes the surface-level morphology of, the 3D version enables retention of the full volumetric continuity to represent the true evolution of the microstructures. Due to the increased complexity and computational cost of the 3D voxel data, the dataset used was limited to 50 images per composition, producing 100 joint samples in total. In the 2D models, 100 images of dimension $128\times128\times128$ per composition were used to train and evaluate the model. 

\begin{figure}[H]
\centering
\begin{minipage}{0.48\textwidth}
    \centering
    \includegraphics[width=\linewidth]{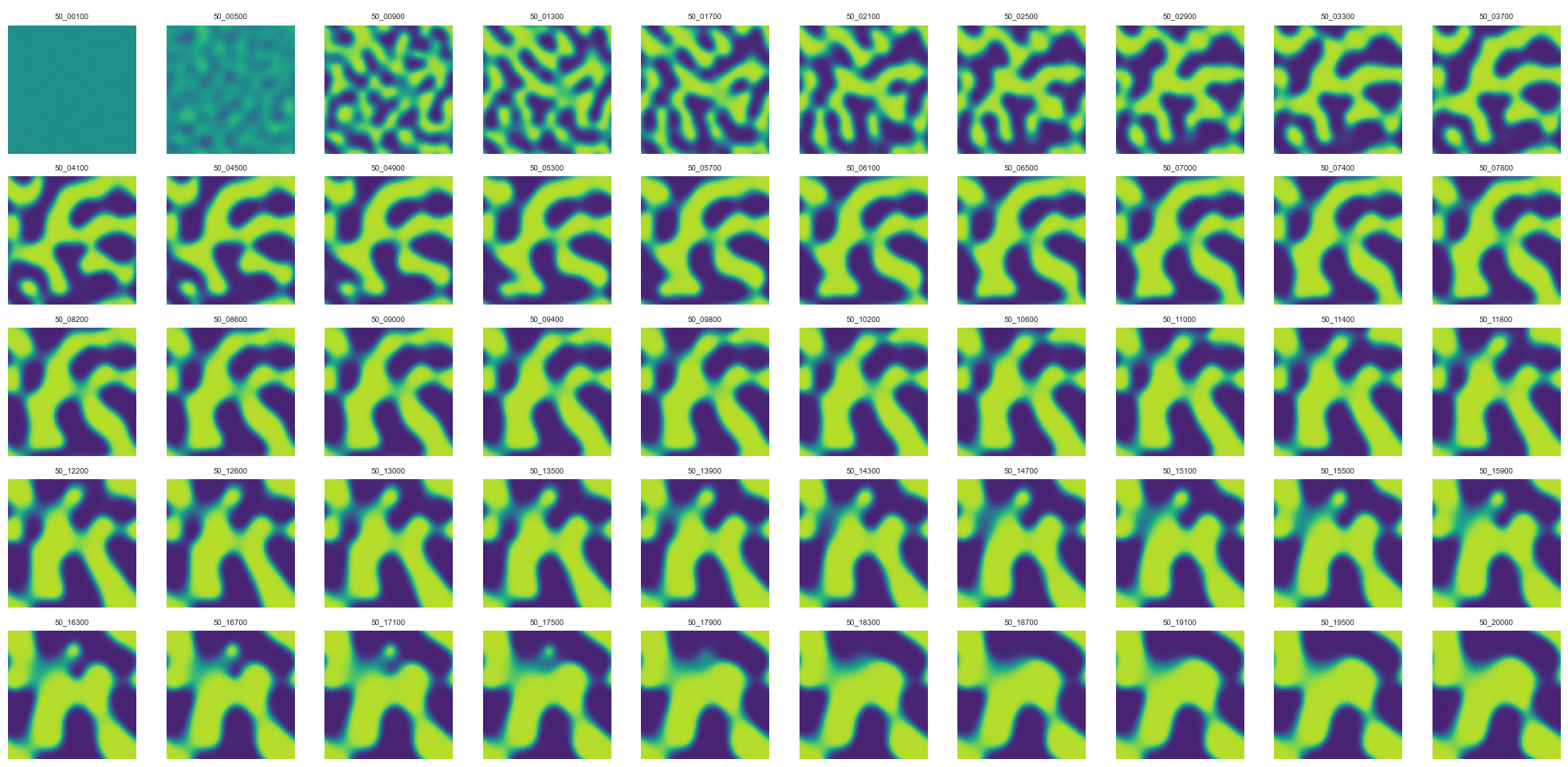}
\end{minipage}
\hfill
\begin{minipage}{0.48\textwidth}
    \centering
    \includegraphics[width=\linewidth]{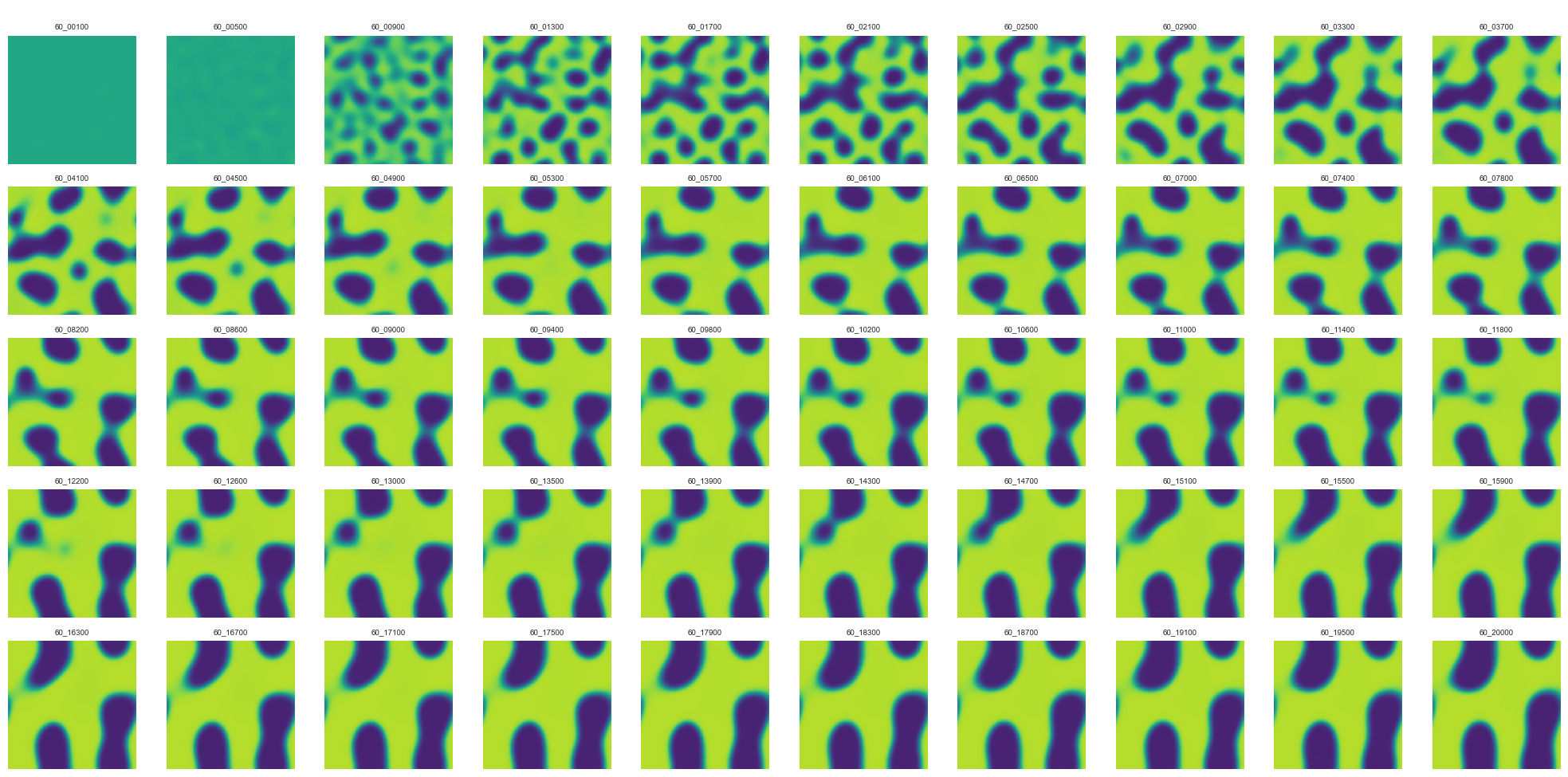}
\end{minipage}
\caption{Input microstructure images of size $128 \times 128 \times 128$ extracted from 
phase-field simulations at 350 K with initial Sb compositions of 0.5 (left) and 
0.6 (right). Each dataset contains 20,000 timesteps for training, while the range 
from 20,000--40,000 timesteps is reserved as unseen data to evaluate long-horizon 
forecasting.}
\label{fig:combined}
\end{figure}

\begin{figure}[H]
\centering
\includegraphics[width=0.70\textwidth]{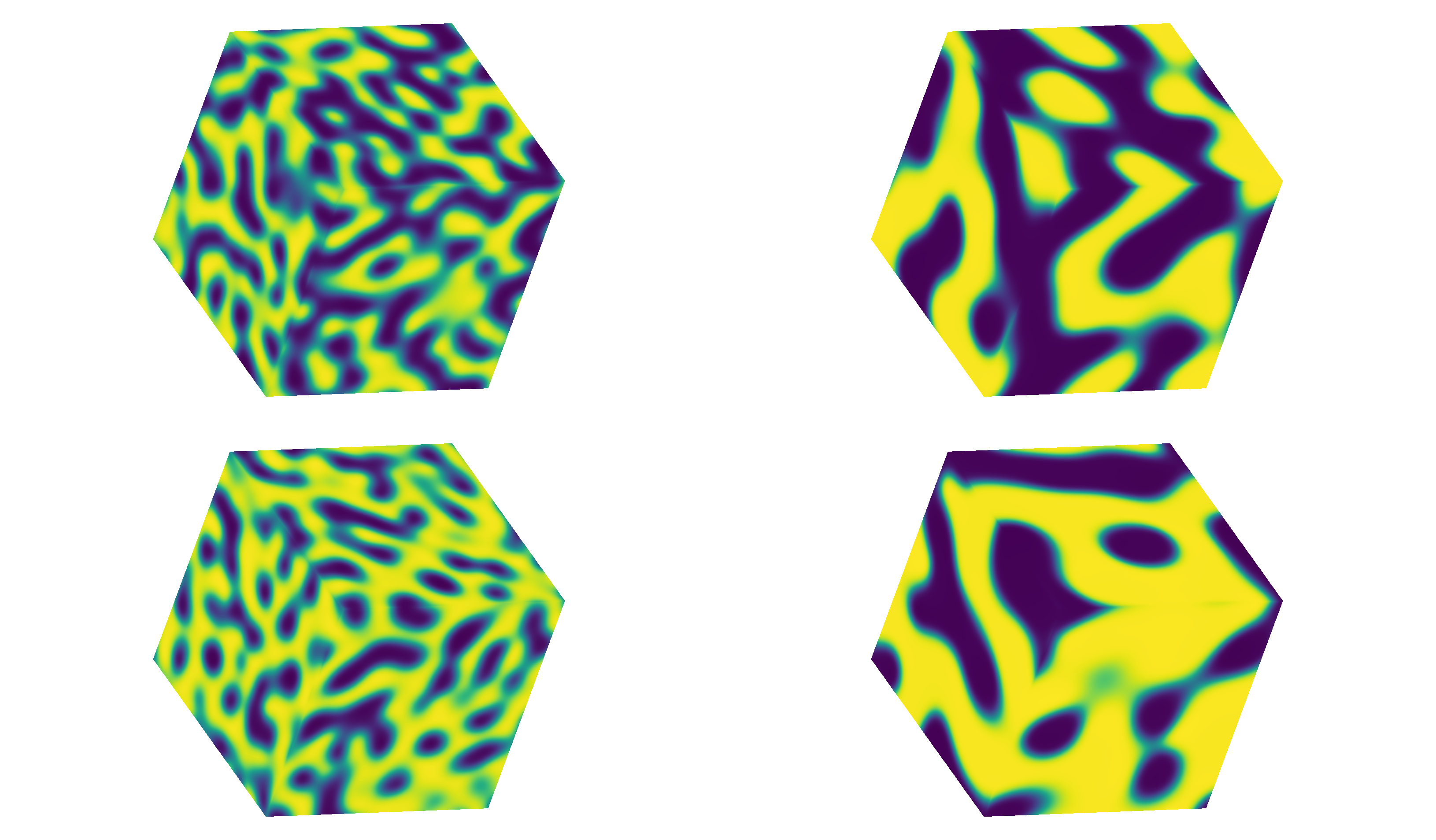}
\caption{Randomly selected 3D voxelized microstructure samples from the joint dataset of 100 images. 
The two samples on the left correspond to an Sb composition of 0.6, while the two on the right correspond to an Sb composition of 0.5.}
\label{fig: single}
\end{figure}

In this 3D model, native volumetric voxel data is preserved without flattening, and each data entry in the volumetric sequence has a shape $(L,2,D,H,W)$. $D$ represents the depth dimension and $2$ represents the total number of channels. The compression in this model is done via a 3D convolutional autoencoder. This architecture is designed to process microstructure inputs of size $(B,2,128,128,128)$, where the two channels correspond to the microstructure field and the composition map. The encoder part is comprised of two blocks, where the first 3D convolution has 32 channels, kernel size 4 with stride 2, and reduces the resolution from $128^3$ to $64^3$. The second convolutional block has 64 latent channels, a kernel size of 4, and a stride of 1, and maintains the $64^3$ resolution. An average $2\times 2\times 2$ pooling layer produces a latent graph representation of $32^3$. The decoder part of the architecture mirrors the encoder part with two transposed convolutional blocks. The first layer expands the latent representation back to 32 channels with a resolution of $64^3$ and introduces a skip connection. The following transposed decoding blocks progressively reconstruct the full resolution back to the $128^3$ output with intermediate $3\times 3\times 3$ convolutions. Finally, a one-channel reconstruction is passed through a sigmoid activation with mean-conservation 
explicitly enforced on the microstructure channel. Overall, the 3D convolutional autoencoder compresses $128^3$ inputs into a $64$-channel latent tensor of size $64^3$, integrates skip connections to preserve fine-scale features, and implements a graph latent head at $32^3$ resolution for graph-based learning. Training and evaluation of the 3D convolutional autoencoder was performed in 80 epochs with a learning rate of \(1\times 10^{-4}\) , with the details of the last epoch mentioned in the table below, along with the convergence visualizations. 

 \begin{table}[h]
\caption{Validation metrics at 80th epoch for $\bar{X}_{\mathrm{Sb}} = 0.5$ and $\bar{X}_{\mathrm{Sb}} = 0.6$ the 3D Convolutional Autoencoder model with input size $128\times128\times128$}\label{tab:sequence_metrics}
\begin{tabular*}{\textwidth}{@{\extracolsep\fill}lcccccc}
\toprule%
& \multicolumn{3}{@{}c@{}}{\textbf{$\mathbf{\bar{X}_{\mathrm{Sb}}} = \mathbf{0.5}$}} & \multicolumn{3}{@{}c@{}}{\textbf{$\mathbf{\bar{X}_{\mathrm{Sb}}} = \mathbf{0.6}$}} \\
\cmidrule{2-4}\cmidrule{5-7}
\textbf{Seq. Length} & \textbf{Valid. MSE} & \textbf{Valid. SSIM} &  
& \textbf{Valid. MSE} & \textbf{Valid. SSIM} & \\
\midrule
3  &0.000250 & 0.997690 & & 0.000225 & 0.997556 & \\
 \\
\botrule
\end{tabular*}
\end{table}

\begin{figure}[H]
\centering
\includegraphics[width=0.9\textwidth]{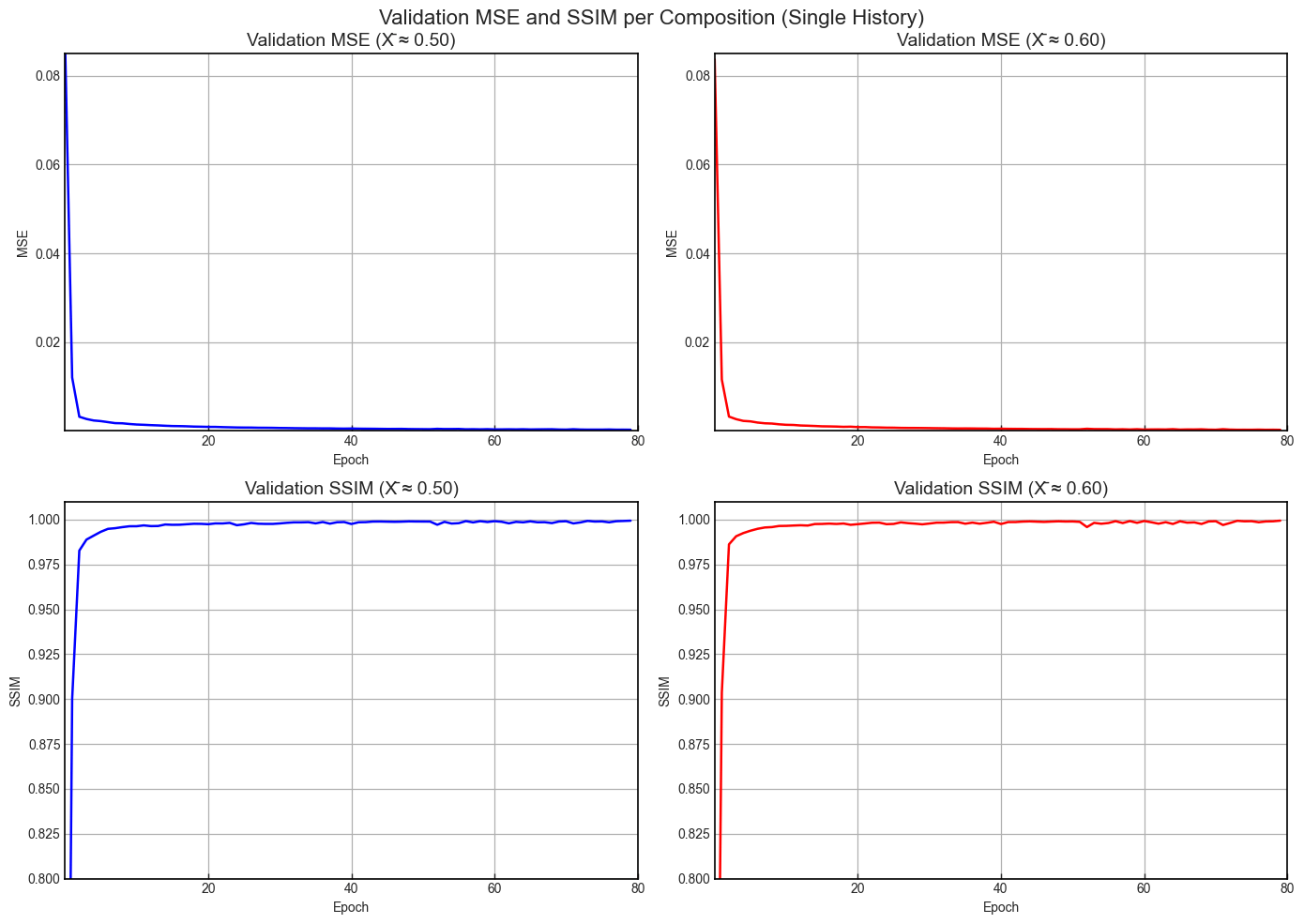}
\caption{Validation of the 3D convolutional autoencoder with $L=3$ and input size 
$128^3$. MSE (top) and SSIM (bottom) over 100 epochs for 
$\bar{X}_{\mathrm{Sb}}=0.5$ (left, blue) and $\bar{X}_{\mathrm{Sb}}=0.6$ 
(right, red) show rapid convergence,
demonstrating robust reconstruction.}
\label{fig4}
\end{figure}

\begin{figure}[H]
\centering
\includegraphics[width=0.7\textwidth]{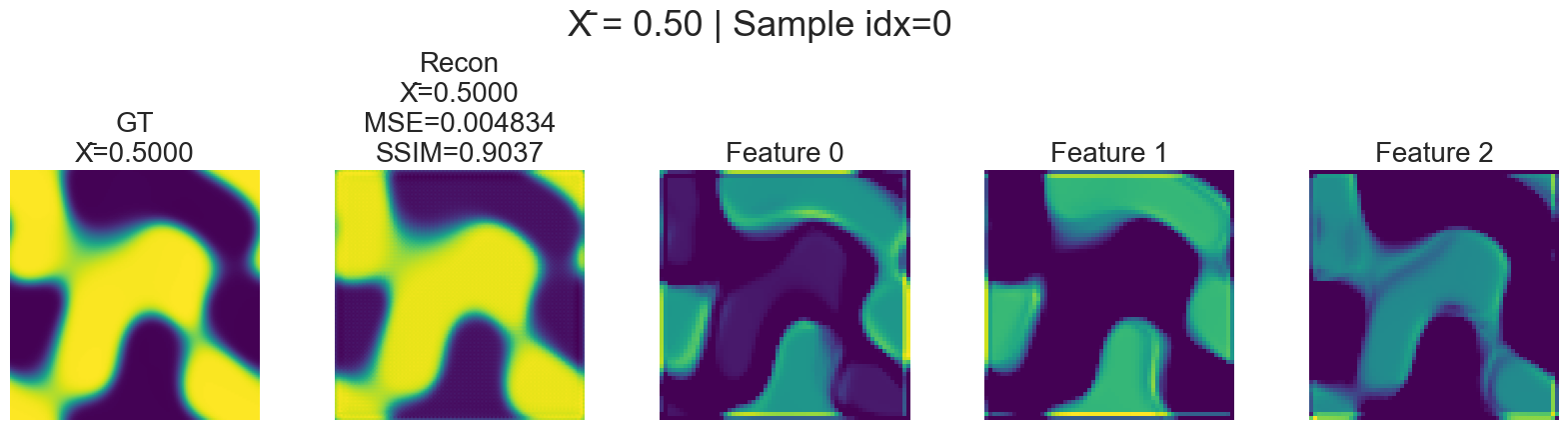}\\[0.5em]
\includegraphics[width=0.7\textwidth]{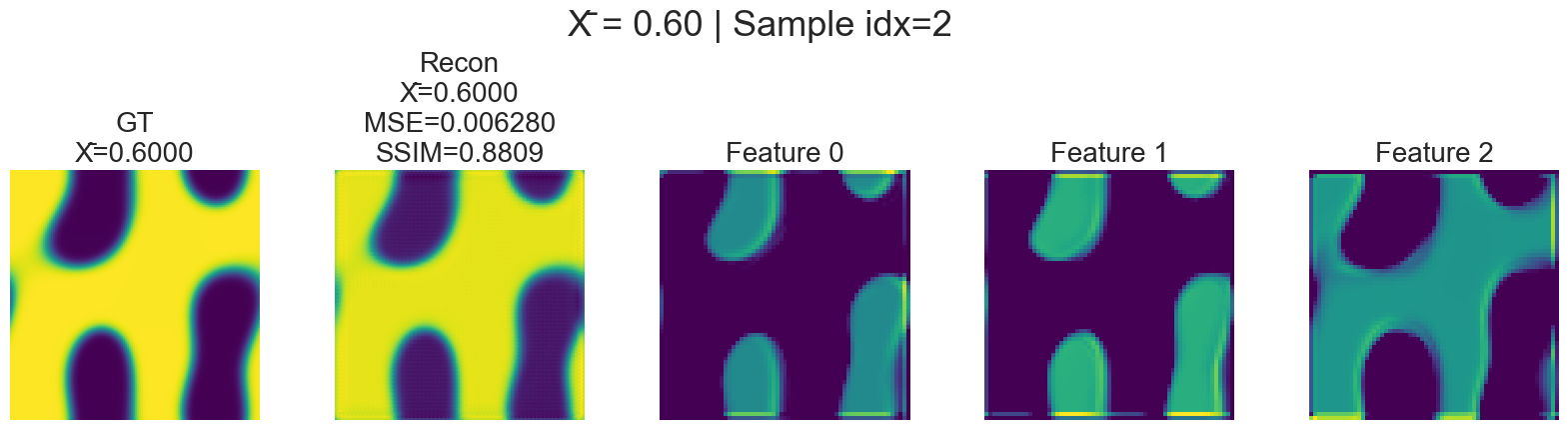}
\caption{3D autoencoder reconstruction results for input size $128 \times 128\times128$ with sequence 
length $L=3$ for $\bar{X}_{\mathrm{Sb}} = 0.5$ (top) and $\bar{X}_{\mathrm{Sb}} = 0.6$ (bottom). 
For each composition, the first column shows the original microstructure (GT), the 
second column is the reconstructed output (Recon), followed by three representative latent 
filters. The reconstructed microstructures are evaluated using SSIM and MSE metrics}
\label{fig:ae_recon_combined}
\end{figure}
In Figure 16, the 3D representations of the reconstructed images are shown next to their corresponding original images. 
\begin{figure}[H]
\centering
\includegraphics[width=0.70\textwidth]{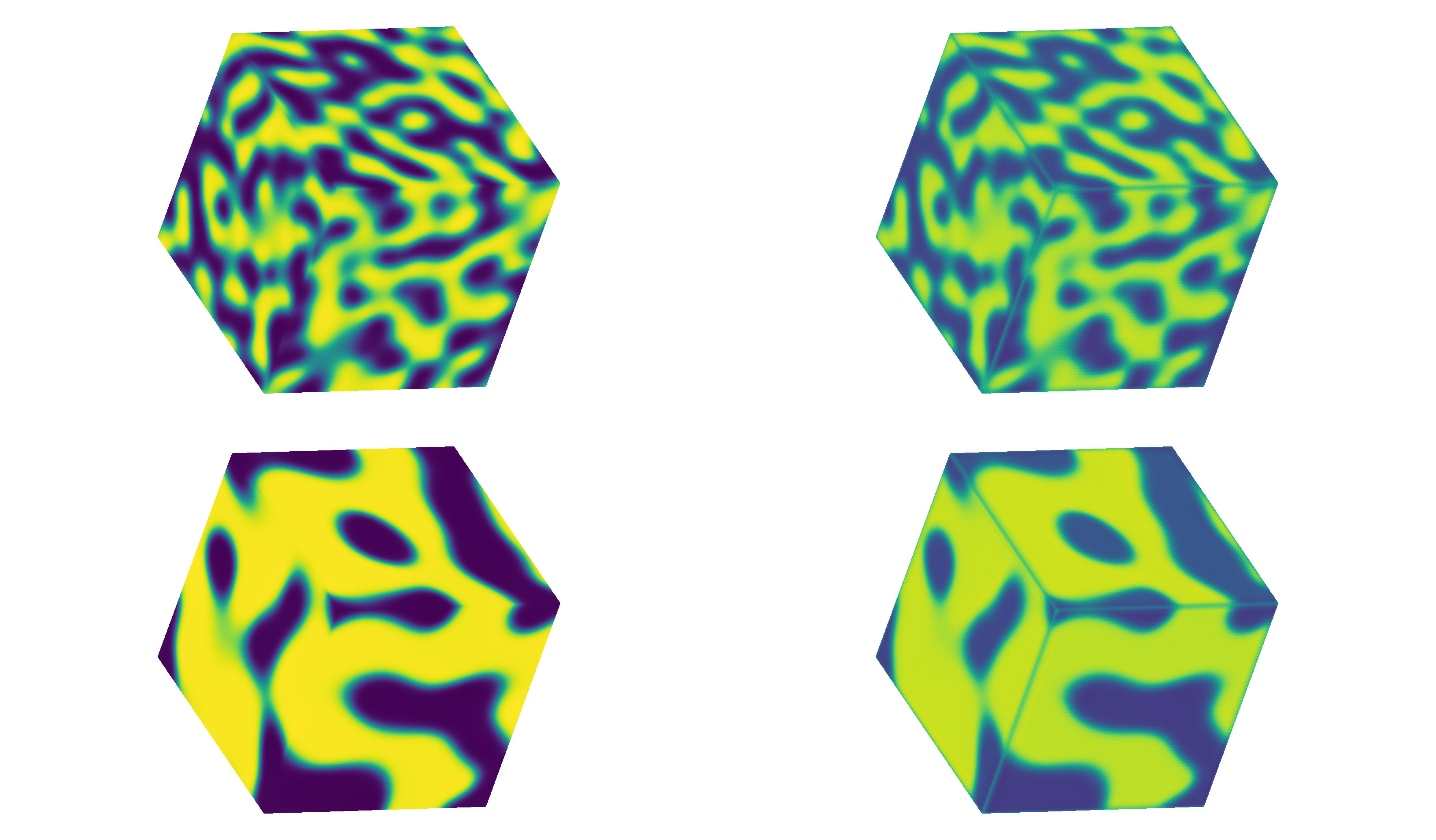}
\caption{Autoencoder reconstruction results for $L = 3$ at $\bar{X}_{\mathrm{Sb}} = 0.5$ (top) and $\bar{X}_{\mathrm{Sb}} = 0.6$ (bottom), showing both output sequences and encoded representations. The left-side images are the original dataset images, and the 3D convolutional autoencoder reconstructs the images on the right side.}
\label{fig:single}
\end{figure}
To convert latents into graphs, a voxel grid of size $(D\times H\times W)$ is converted into an undirected 6-neighborhood graph by connecting adjacent voxels along depth, height, and width. In the next step, latent tensors are flattened and at each timestep assembled into torch geometric graphs. For inspection, the ground-truth microstructure is resampled to the latent resolution and rendered in Napari, with latent nodes overlaid as value-colored points that can be seen in the figure below.

\begin{figure}[H]
\centering
\includegraphics[width=1.0\textwidth]{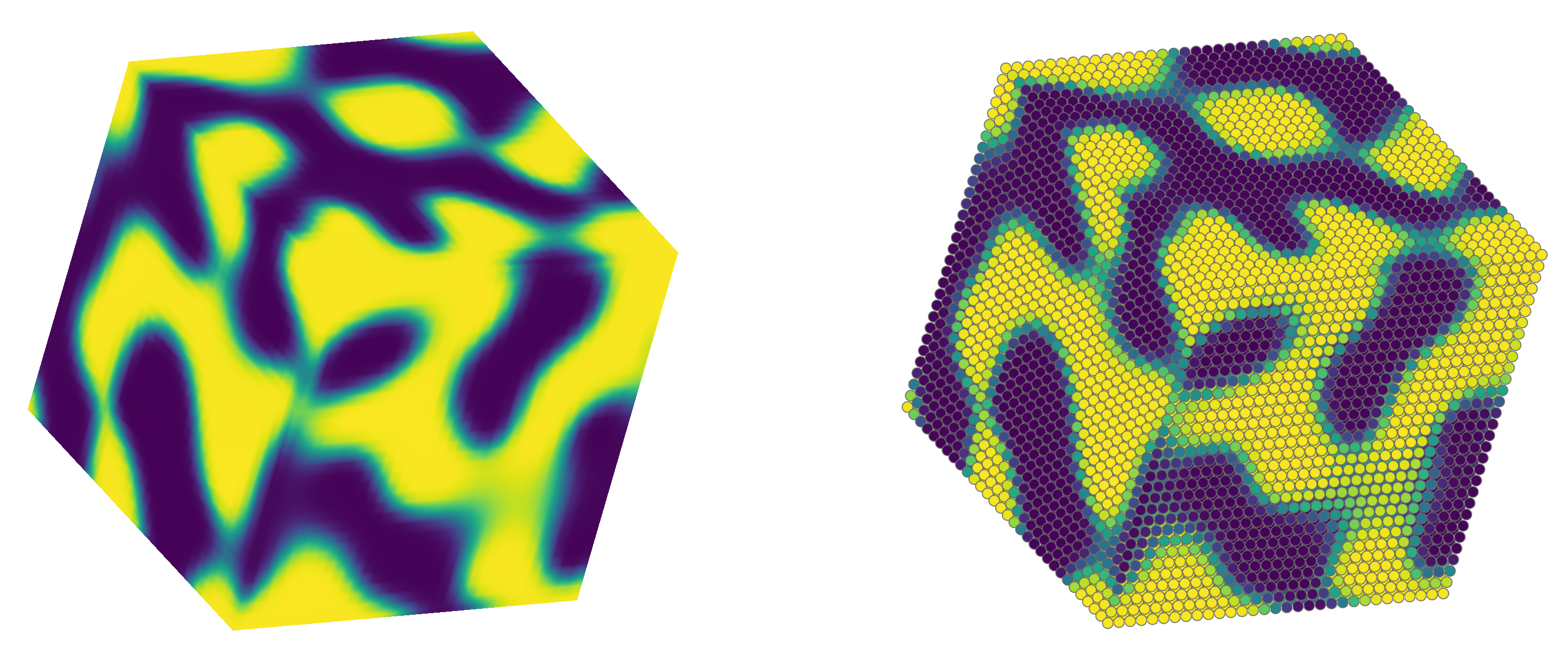}
\caption{Latent graph overlaid on the resampled 3D microstructure at composition $\bar{X}_{\mathrm{Sb}}=0.5$: nodes on the $D\times H\times W$ lattice visualize the encoded features, enabling direct spatial comparison between the volumetric field and its graph representation.}
\label{fig:single}
\end{figure}

The GCN-LSTM-PI architecture operates on a fixed voxel-lattice graph with $N$ = $D\times H\times W$ nodes per time step and node features of dimension $65$ (concatenating $64$ latent channels with scalar composition). This architecture uses a stack of three  \texttt{GCNConv} layers (width 320) with per-node LayerNorm, ReLU, and $0.2$ dropout, with residual connections that are applied at every block, with a learned input projection $65\!\to\!320$ in the first block. To capture temporal dynamics, the embeddings per node in the $T$ time steps are fed to an LSTM (hidden size 320), while the last hidden state is LayerNorm, dropped out, and linearly mapped to 64 output channels per node, producing a prediction tensor reshaped to $(B,64,D,H,W)$. The LSTM is applied per node (the sequence tensor is reshaped to \((B\times N,\, T,\, 320)\), and only the first GCN block uses the \(65\!\to\!320\) linear residual projection; subsequent GCN blocks use identity skip connections. With this setup, graph convolutions capture the local spatial structures in the shared $D\times H\times W$ topology, while the LSTM accounts for the temporal evolution of the microstructures. 

\subsection{Use of Large Language Models}\label{subsec9}

In this work, a local large language model, and OpenAI models (models GPT\mbox{-}4o and GPT\mbox{-}5 Thinking) were used for coding assistance and minor editorial tasks (e.g., figure captions). All outputs were reviewed and validated by the author. 

\section{Results}\label{sec5}

\subsection{Training and Evaluating the Model }\label{subsec1}

The first result examined is the architecture of the 2D convolutional autoencoder with input images of size 128$\times$128. Before defining the training function, a mid-epoch visualization is implemented for both compositions (\( \bar{X}_{\mathrm{Sb}} = 0.5 \) and \( 0.6 \)) over various sequence lengths. This function enables the analysis of real-time comparisons between the predicted and original microstructure images during training, allowing the model hyperparameters to be adjusted if the visualizations and evaluation metrics are not optimal. This visualization is performed for the evaluation batch every 50 epochs and is based on the latent graph sequences trained by the GCN-LSTM-PI model and decoded by the convolutional 2D autoencoder. Conservation of mean composition is enforced on the decoded predictions, and evaluation metrics, including mean squared error (MSE), structural similarity index (SSIM), and predicted vs. true mean \( \bar{X}_{\mathrm{Sb}} \) values are computed and visualized for each subplot.     

The training method for our model is performed jointly for both compositions, $\bar{X} = 0.5$ and $\bar{X} = 0.6$, as they are combined and fed simultaneously into the model during both training and evaluation. The batches are sampled from the \texttt{train\_loader} or \texttt{val\_loader} and comprise both compositions. In this setup, the GCN-LSTM-PI model and the autoencoder learn a unified representation that generalizes across varying composition values, allowing us to adjust this model in the future to accommodate more than just two varying compositions. The GCN-LSTM-PI pipeline processes the input graph sequences to predict the next latent frame, and these latent states are then decoded via the 2D convolutional autoencoder to reconstruct the predicted microstructure images. During the training process, one \textbf{Total Loss} is calculated over the full batch, and is composed of five weighted components: \textbf{image reconstruction loss (MSE)}, \textbf{structural similarity loss (SSIM)}, \textbf{composition conservation loss}, \textbf{physics-informed loss} based on the Cahn-Hilliard equation, and \textbf{latent loss} between predicted and true encodings. MSE and SSIM, and physics losses are computed on the microstructure channel, while the latent loss is computed between predicted and true latent features. Since we have defined the functions for all other losses before, below is the function of the latent loss:

\[
\mathcal{L}_{\text{latent}}
= \frac{1}{B} \sum_{i=1}^{B} \frac{1}{N}
  \big\| z^{\mathrm{pred}}_i - z^{\mathrm{true}}_i \big\|_F^2.
\]
In the latent loss, \(B\) is the batch size, \(z_i^{\mathrm{pred}}\) refers to the predicted latents of GCN-LSTM-PI for the sample \(i\), \(z_i^{\mathrm{true}}\) is the latent target of the encoder, \(N\) is the number of elements per latent (2D: \(N=CHW\), 3D: \(N=CDHW\))

The advantage of joint training over separately training sequential models is improved generalization. Upon performing separate sequential training models for both compositions, it became apparent that the model is biased, favoring the last composition that was trained on. Joint training encourages the model to learn the spatial and temporal features that are commonly shared between varying compositions, which is valuable when the structures share a similar underlying physics but are composition-dependent. It also eliminates the need to train, validate, and maintain multiple models, as only one model is used in the joint training, which significantly helps save computational resources and simplifies the inference pipeline. The joint training function enables the model to process a graph sequence generated from the compressed latent tensors, decode the latent predicted frame, and apply the mean conservation to match the original image's mean composition. At the end of each epoch, the defined loss function is used to perform validation and compute metrics that are logged globally and per composition. The checkpoints are saved on the computer disk and contain the model, decoder, epoch number, and history of the training and validation. At the end of each epoch, validation is performed using the same loss formulation. Metrics are calculated and recorded globally and per composition. The checkpoints, which contain the model, decoder, epoch number, and history dictionary, are saved to disk if a save directory is provided. Mid-epoch visualizations are triggered every 50 epochs, offering qualitative insights into temporal prediction fidelity and composition-specific performance. This comprehensive training loop enables optimization for numerical accuracy, physical consistency,  composition conservation, and generalization across multiple sequence lengths and alloy compositions. The pre-trained autoencoder is fine-tuned end-to-end jointly with the GCN-LSTM-PI model. The model is trained for 100 epochs using the following hyperparameters, which were modified and adjusted to obtain the desired results with the following hyperparameters:
\begin{itemize}
    \item Learning rate: \( 10^{-4} \)
    \item Image MSE loss weight: \( \lambda_{\text{img}} = 1.0 \)
    \item SSIM loss weight: \( \lambda_{\text{ssim}} = 1.0 \)
    \item Conservation loss weight: \( \lambda_{\text{cons}} = 0.03 \)
    \item Physics-informed loss weight: \( \lambda_{\text{phys}} = 0.05 \)
    \item Latent loss weight: \( \lambda_{\text{latent}} = 3.0 \)
\end{itemize}

Each training run produces a distinct history object that tracks all loss components and evaluation metrics. All models and metrics are stored in composition-specific subdirectories under a central log directory. This strategy enables the GCN-LSTM-PI model to learn across diverse temporal dynamics and alloy compositions while preserving physical interpretability and compositional consistency.

\begin{figure}[H]
\centering
\includegraphics[width=1.0\textwidth]{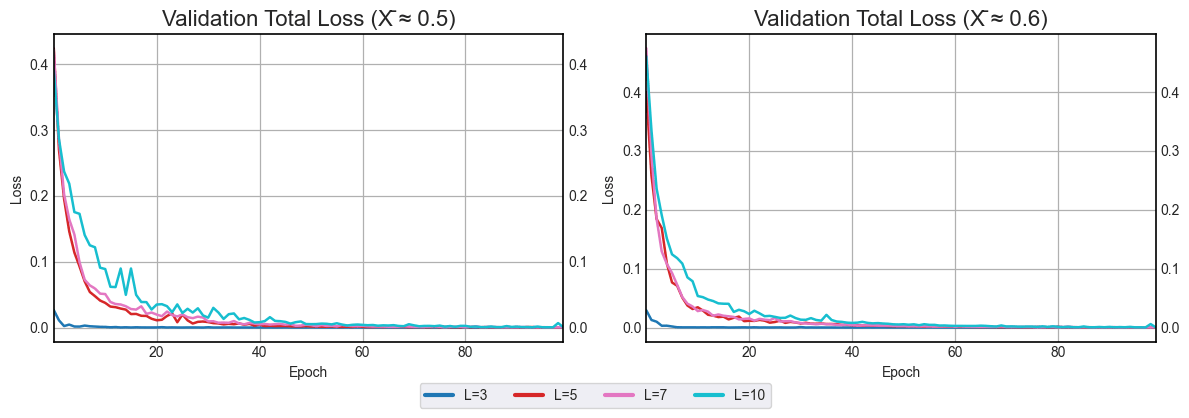}
\caption{Validation total loss over 100 epochs for the 2D GCN-LSTM-PI model with input size $128 \times 128$ and sequence lengths $L = 3, 5, 7, 10$ at 
$\bar{X}_{\mathrm{Sb}} = 0.5$ (left) and $\bar{X}_{\mathrm{Sb}} = 0.6$ (right). The total 
loss combines SSIM, MSE, physics-informed, conservation, and latent losses. All 
sequences converge rapidly, with longer $L$ showing minor early fluctuations but 
similar final values near zero.}
\label{fig5}
\end{figure}

The total validation loss curves in Figure 18 indicate stable convergence across all sequence lengths for both compositions $\bar{X}_{\mathrm{Sb}} = 0.5$ and $\bar{X}_{\mathrm{Sb}} = 0.6$. Among the four sequence lengths, the model trained with $L = 3$ consistently exhibits the fastest initial convergence, often reaching convergence within the first few epochs. To further display the quantification of the model performance, Tables 4 and 5 were used to compare the final validation metrics at epoch 100 for all loss metrics. The sequence length $L = 3$ shows the best performance in terms of \texttt{Total Loss}, \texttt{MSE}, \texttt{SSIM}, and \texttt{Latent Loss} for both compositions. The only metric where $L = 3$ did not perform the best was the \texttt{Conservation Loss}; however, the values remained extremely small (on the order of $10^{-15}$), due to the conservation of the mean composition enforced during training. Across all sequence lengths, the \texttt{SSIM} remains remarkably high, approaching 1.0, indicating excellent structural similarity between predicted and ground-truth microstructure images.

\begin{table}[h]
\centering
\caption{Validation losses at epoch 100 of \textbf{$\bar{X}_{\mathrm{Sb}} = 0.5$} across different sequence lengths for the 2D
GCN-LSTM-PI model with input size $128\times128$}
\label{tab:x0.6_losses}
\begin{tabular*}{\textwidth}{@{\extracolsep\fill}lcccccc}
\toprule
\textbf{Seq. Length} & \textbf{Total Loss} & \textbf{MSE} & \textbf{Physics Loss} & \textbf{Latent Loss} & \textbf{Conse. Loss} & \textbf{SSIM} \\
\midrule
3  & 0.000039 & 0.00000142 & 0.00000363 & 0.000004 & 0.0000000000 & 1.0000 \\
5  & 0.000528 & 0.00001157 & 0.00003003 & 0.000109 & 0.0000000000 & 0.9998 \\
7  & 0.000738 & 0.00002879 & 0.00007418 & 0.000074 & 0.0000000000 & 0.9995 \\
10 & 0.000638 & 0.00001579 & 0.00004041 & 0.000162 & 0.0000000000 & 0.9999 \\
\botrule
\end{tabular*}
\end{table}

\begin{table}[h]
\centering
\caption{Validation losses at epoch 100 of \textbf{$\bar{X}_{\mathrm{Sb}} = 0.6$} across different sequence lengths for the 2D
GCN-LSTM-PI model with input size $128\times128$}
\label{tab:x0.6_losses}
\begin{tabular*}{\textwidth}{@{\extracolsep\fill}lcccccc}
\toprule
\textbf{Seq. Length} & \textbf{Total Loss} & \textbf{MSE} & \textbf{Physics Loss} & \textbf{Latent Loss} & \textbf{Conse. Loss} & \textbf{SSIM} \\
\midrule
3  & 0.000054 & 0.00000185 & 0.00000392 & 0.000004 & 0.0000000000 & 1.0000 \\
5  & 0.000333 & 0.00001389 & 0.00002890 & 0.000050 & 0.0000000000 & 0.9998 \\
7  & 0.000772 & 0.00002887 & 0.00005940 & 0.000081 & 0.0000000000 & 0.9995 \\
10 & 0.000435 & 0.00002003 & 0.00004223 & 0.000099 & 0.0000000000 & 0.9999 \\
\botrule
\end{tabular*}
\end{table}
\subsection{Decoding Predicted Latents to Images}\label{subsec2}

In the training phase, the 2D convolutional autoencoder (both encoder and decoder) is trained jointly with the GCN-LSTM-PI model. Once the GCN-LSTM-PI model receives a sequence of graph-structured latent inputs, it predicts the latent representation of the next timestep once it is fed with the graph sequences. During the same time, the original image of the microstructure is passed through the encoder section of the autoencoder to produce a latent representation with a set of skip connections. These skip connections are implemented to help preserve the spatial information lost during the downsampling operation of the encoder. The decoder then reconstructs the predicted image $\hat{x}_{t+1}$ from the latent $z^{\mathrm{pred}}$ using those skip connections. With this architecture, the decoder leverages the spatial context during the reconstruction process to help improve the fidelity of the predicted microstructure images. The decoder outputs the microstructure channel. Here, composition conservation is also enforced on the microstructure channel to ensure preserving the global composition of the predicted image, since it should be consistent with the original image. Specifically, the decoded image $\hat{x}_{t+1}$ is corrected using:

\begin{equation}
\hat{x}_{t+1} = \hat{x}_{t+1} - \mu_{\text{pred}} + \mu_{\text{true}},
\end{equation}

where $\mu_{\text{pred}}$ and $\mu_{\text{true}}$ denote the mean pixel intensity of the predicted and ground-truth microstructure channels, respectively. This correction ensures strict preservation of the mean composition during both training and inference.

\subsection{Visualization of Predictions}\label{subsec3}

To facilitate comparisons between the original and predicted microstructure images for compositions $\bar{X}_{\mathrm{Sb}} = 0.5$ and $\bar{X}_{\mathrm{Sb}} = 0.6$, a visualization is performed and shown in Figure 19 for the four sequence lengths. For consistency, the same set of three validation samples is selected for each sequence length based on closely matching composition values. For each sample $L$, the GCN-LSTM-PI model predicts the latent representation of the next timestep. This latent representation is decoded into a predicted image using the corresponding trained autoencoder. The predicted and ground-truth microstructure channels are plotted side by side, along with their mean composition ($\bar{X}$), mean squared error MSE, and SSIM score. This visualization helps us with an assessment of how model performance evolves with different input sequence lengths. It also helps validate both accuracy and conservation across predictions. 

%\begin{table}[h]
%\caption{Visualization of GCN-LSTM-PI Validation metrics for $\math{\bar{X}_{\mathrm{Sb}}} = \math{0.5}$ and $\math{0.6}$ for the 2D
%Convolutional Autoencoder model with input size $128\times128$}
\label{tab:sequence_metrics}
%\begin{tabular*}{\textwidth}{@{\extracolsep\fill}lcccccc}
%\toprule
%& \multicolumn{3}{@{}c@{}}{\textbf{$\mathbf{\bar{X}_{\mathrm{Sb}}} = \mathbf{0.5}$}} & \multicolumn{3}{@{}c@{}}{\textbf{$\mathbf{\bar{X}_{\mathrm{Sb}}} = \mathbf{0.6}$}} \\
%\cmidrule{2-4}\cmidrule{5-7}
%\textbf{Seq. Length} & \textbf{MSE} & \textbf{SSIM} & \textbf{Composition} & \textbf{MSE} & \textbf{SSIM} & \textbf{Composition} \\
%\midrule
%3  & 0.000002 & 0.9999 & 0.50 & 0.000002 & 0.9999 & 0.60 \\
%5  & 0.000053 & 0.9989 & 0.50 & 0.000013 & 0.9996 & 0.60 \\
%7  & 0.000044 & 0.9989 & 0.50 & 0.000033 & 0.9988 & 0.60 \\
%10 & 0.000042 & 0.9990 & 0.50 & 0.000032 & 0.9993 & 0.60 \\
%\botrule
%\end{tabular*}
%\end{table}
This visualization provides an intuitive overview of how predictive performance varies with sequence length while also highlighting the consistency of predictions for composition conservation.
\begin{figure}[H]
\centering
\includegraphics[width=1\textwidth]{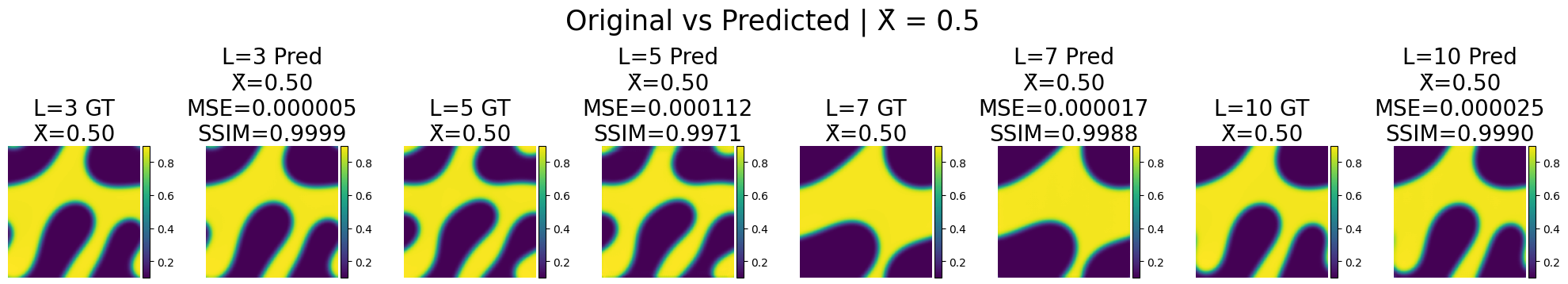}
\includegraphics[width=1\textwidth]{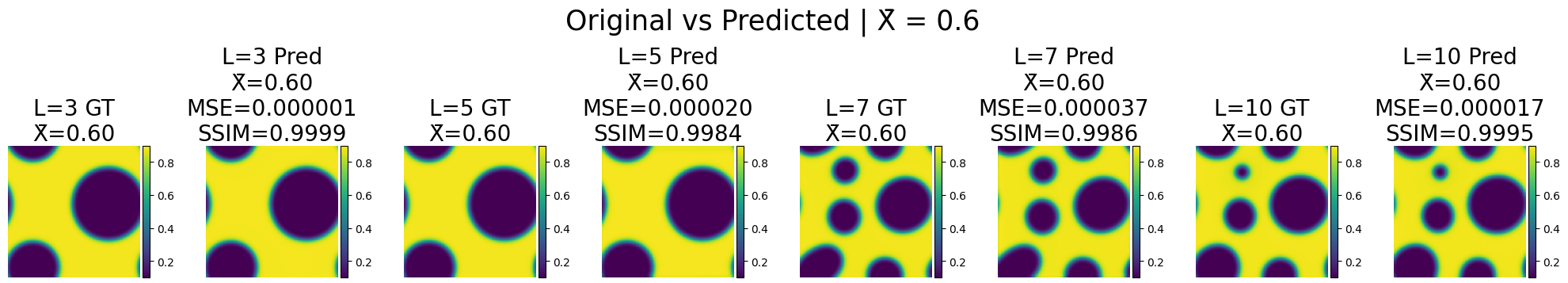}
\caption{Ground Truth  (GT) and predicted microstructures (Pred) for $\bar{X}_{\mathrm{Sb}}=0.5$ (top) and $0.6$ 
(bottom) at $L=3,5,7,10$. Next-step predictions decoded from GCN-LSTM latents are 
shown with mean composition, MSE, and SSIM, illustrating accuracy and conservation 
for the 2D $128 \times 128$ model.}
\label{fig:ae_recon_combined}
\end{figure}

\subsection{Predicting Long-Horizon Microstructures }\label{subsec3}

One of the main advantages of this method relative to phase field simulation is the speed with which future forecasting is performed. After completion of the model training, future forecasting happens very quickly in just under one minute, which is considerably faster than running the entire phase-field simulation. To evaluate the model’s ability to generalize beyond the training distribution, forecasting is performed on a held-out set of unseen microstructure data. The dataset was originally generated using Cahn–Hilliard simulations up to $t = 200{,}000$ timesteps for each composition. The first half of the simulation data is used (from $t = 0$ to $t = 100{,}000$) for training and validation, and the second half ($t = 100{,}000$ to $t = 200{,}000$) is reserved as an unseen test set to evaluate the forecast ability of the model. 

Long-horizon forecasting is based on the one-shot forecasting strategy. The model first predicts the next latent state from the context once and then decodes it. The single prediction is compared against multiple future ground-truth frames to assess the quality of the extrapolation. 

\begin{figure}[H]
\centering
\begin{minipage}{\textwidth}
  \centering
  \includegraphics[width=0.5\textwidth]{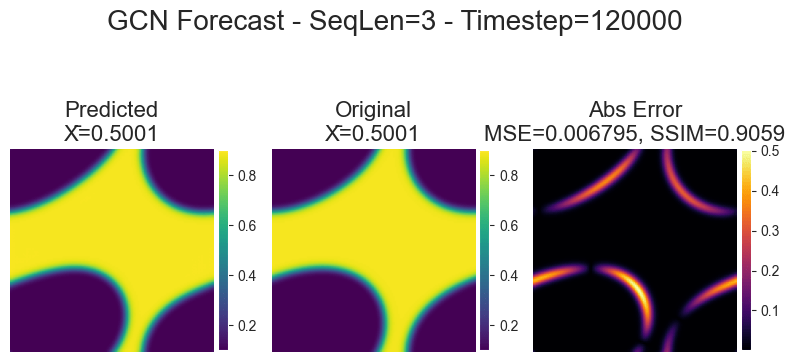}\par\vspace{0.5em}
  \includegraphics[width=0.5\textwidth]{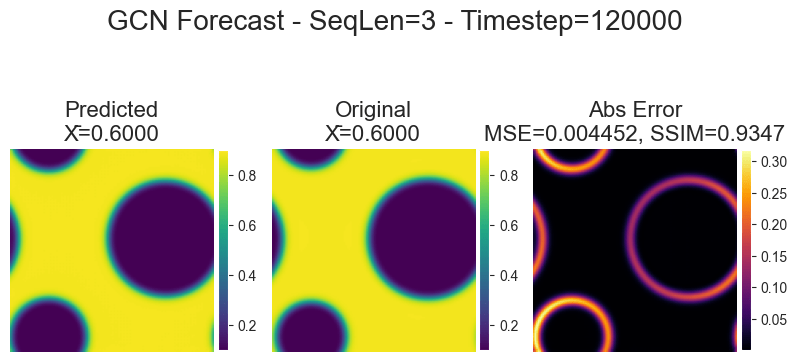}
\end{minipage}
\caption{GCN-LSTM-PI forecast results at $\bar{X}_{\mathrm{Sb}}=0.5$ (top) and $\bar{X}_{\mathrm{Sb}}=0.6$ (bottom) for $L=3$. Each row shows the predicted microstructure, ground truth, and error map for future timesteps 20{,}000 steps beyond the training range (trained up to timestep 100{,}000) for the 2D $128\times128$ model.}
\label{fig:future_forecast_gcn}
\end{figure}
\begin{table}[h]
\caption{Long-horizon forecast metrics for sequence length $L = 3$ at future timesteps 
for $\mathbf{\bar{X}_{\mathrm{Sb}}} = 0.5$ and $0.6$ of the 2D model with input size $128\times128$}\label{tab:forecast_metrics}
\begin{tabular*}{\textwidth}{@{\extracolsep\fill}lcccccccc}
\toprule
& \multicolumn{4}{@{}c@{}}{$\mathbf{\bar{X}_{\mathrm{Sb}}} = 0.5$} 
& \multicolumn{4}{@{}c@{}}{$\mathbf{\bar{X}_{\mathrm{Sb}}} = 0.6$} \\
\cmidrule{2-5}\cmidrule{6-9}
\textbf{Timestep} & \textbf{MSE} & \textbf{SSIM} & $\mathbf{\bar{X}_{\mathrm{pred}}}$ & $\mathbf{\bar{X}_{\mathrm{true}}}$ 
& \textbf{MSE} & \textbf{SSIM} & $\mathbf{\bar{X}_{\mathrm{pred}}}$ & $\mathbf{\bar{X}_{\mathrm{true}}}$ \\
\midrule
100500  & 0.000048 & 0.9990 & 0.5001 & 0.5001 & 0.000019 & 0.9996 & 0.6000 & 0.6000 \\
101000  & 0.000085 & 0.9982 & 0.5001 & 0.5001 & 0.000033 & 0.9993 & 0.6000 & 0.6000 \\
110000  & 0.002258 & 0.9610 & 0.5001 & 0.5001 & 0.001093 & 0.9813 & 0.6000 & 0.6000 \\
120000  & 0.006796 & 0.9059 & 0.5001 & 0.5001 & 0.004452 & 0.9347 & 0.6000 & 0.6000 \\
130000  & 0.012194 & 0.8592 & 0.5001 & 0.5001 & 0.010698 & 0.8722 & 0.6000 & 0.6000 \\
140000  & 0.017669 & 0.8230 & 0.5001 & 0.5001 & 0.020336 & 0.8066 & 0.6000 & 0.6000 \\
150000  & 0.022846 & 0.7950 & 0.5001 & 0.5001 & 0.033732 & 0.7449 & 0.6000 & 0.6000 \\
160000  & 0.027568 & 0.7732 & 0.5001 & 0.5001 & 0.051350 & 0.6885 & 0.6000 & 0.6000 \\
170000  & 0.031791 & 0.7559 & 0.5001 & 0.5001 & 0.074810 & 0.6354 & 0.6000 & 0.6000 \\
180000  & 0.035528 & 0.7420 & 0.5001 & 0.5001 & 0.108862 & 0.5917 & 0.6000 & 0.6000 \\
190000  & 0.038811 & 0.7306 & 0.5001 & 0.5001 & 0.108978 & 0.5908 & 0.6000 & 0.6000 \\
200000  & 0.041687 & 0.7213 & 0.5001 & 0.5001 & 0.108987 & 0.5906 & 0.6000 & 0.6000 \\
\botrule
\end{tabular*}
\end{table}

For forecasting, a context window of $L$ timesteps ending at $t = 100{,}000$ is provided as input. Each microstructure image in this sequence is encoded using a convolutional autoencoder, yielding a sequence of latent feature maps. These latents are then converted into a graph sequence with $4{,}096$ nodes per graph, where each node contains 256 latent channels and one scalar composition value. The graphs use a fixed 4-neighbor spatial connectivity and are passed to a GCN-LSTM model, which predicts the latent representation of the next timestep (i.e., $t = 100{,}500$). This latent prediction is then decoded back into a full-resolution microstructure image using the autoencoder’s decoder, with a projection step of mean composition to enforce conservation.

Although only a single future latent state is predicted, the resulting prediction is evaluated against a sequence of unseen ground-truth microstructure images from $t = 100{,}500$ to $t = 200{,}000$, in 500-timestep increments. This one-shot direct forecast is assessed over long temporal horizons to quantify how well the model generalizes to future microstructure states. The calculated metrics include mean squared error (MSE), structural similarity index (SSIM), and physics-informed loss based on the Cahn–Hilliard residual to evaluate the physical and structural fidelity of the predictions.

The subsequent results analyzed belong to the 2D architecture with inputs of 256$\times$256 over 100 epochs. The training method for this model replicates the 128×128 setup and differs only in adjusting the loss weights: with more weight emphasized on physics, conservation, and SSIM; less on the latent term (image MSE unchanged). 

\begin{itemize}

  \item Learning rate: \(10^{-4}\)
  \item Image MSE loss weight: \(\lambda_{\text{img}} = 1.0\)
  \item SSIM loss weight: \(\lambda_{\text{ssim}} = 5.0\)
  \item Conservation loss weight: \(\lambda_{\text{cons}} = 0.1\)
  \item Physics-informed loss weight: \(\lambda_{\text{phys}} = 3.0\)
  \item Latent loss weight: \(\lambda_{\text{latent}} = 0.5\)
\end{itemize}

\begin{figure}[H]
\centering
\includegraphics[width=1.0\textwidth]{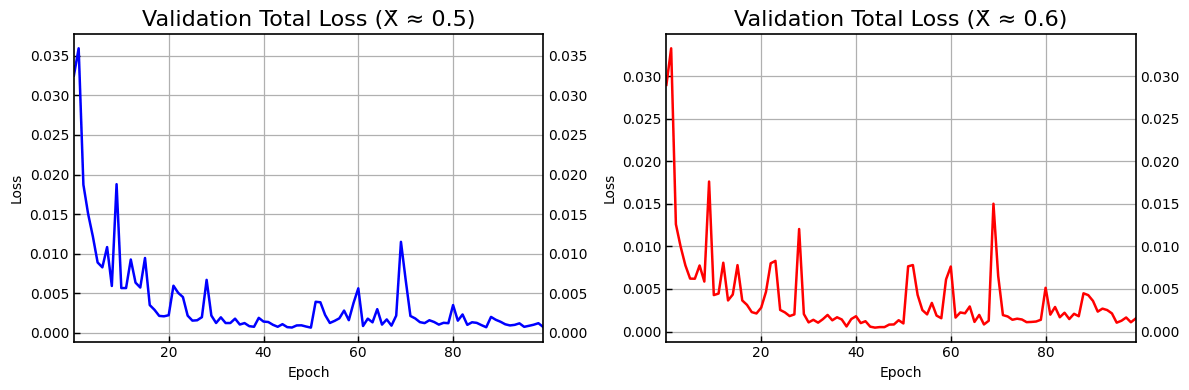}
\caption{Validation total loss over 100 epochs for the 2D GCN-LSTM-PI model with input size $256 \times 256$ at 
$\bar{X}_{\mathrm{Sb}} = 0.5$ (left) and $\bar{X}_{\mathrm{Sb}} = 0.6$ (right). The total 
loss combines SSIM, MSE, physics-informed, conservation, and latent terms. Both 
compositions start to converge after epoch 20 with occasional spikes. }
\label{fig5}
\end{figure}
% Requires \usepackage{booktabs}
\begin{table}[h]
\centering
\caption{Validation losses at epoch 100 of $\bar{X}_{\mathrm{Sb}} = 0.5$ and $\bar{X}_{\mathrm{Sb}} = 0.6$ for the 2D GCN-LSTM-PI model with input size $256\times256$.}
\label{tab:x_losses_combined_epoch99}
\begin{tabular*}{\textwidth}{@{\extracolsep\fill}llcccccc}
\toprule
\textbf{Compos.} & \textbf{Total Loss} & \textbf{MSE} & \textbf{Physics Loss} & \textbf{Latent Loss} & \textbf{Cons.\ Loss} & \textbf{SSIM} \\
\midrule
\(\bar{X}_{\mathrm{Sb}}=0.5\) & 0.000918 & 0.00000827 & 0.00002244 & 0.000021 & 0.0000000000 & 0.9998 \\
\(\bar{X}_{\mathrm{Sb}}=0.6\) & 0.000451 & 0.00000537 & 0.00001147 & 0.000020 & 0.0000000000 & 0.9999 \\
\botrule
\end{tabular*}
\end{table}

\begin{figure}[H]
\centering
\includegraphics[width=1\textwidth]{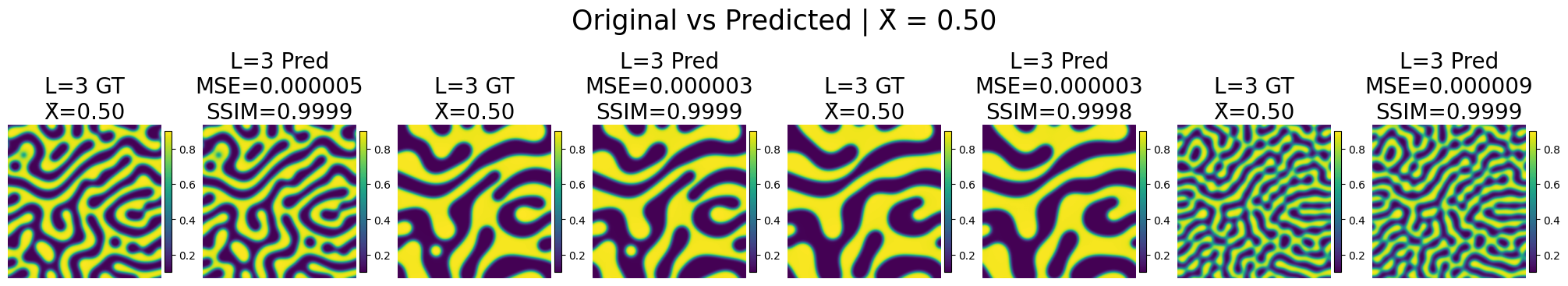}
\includegraphics[width=1\textwidth]{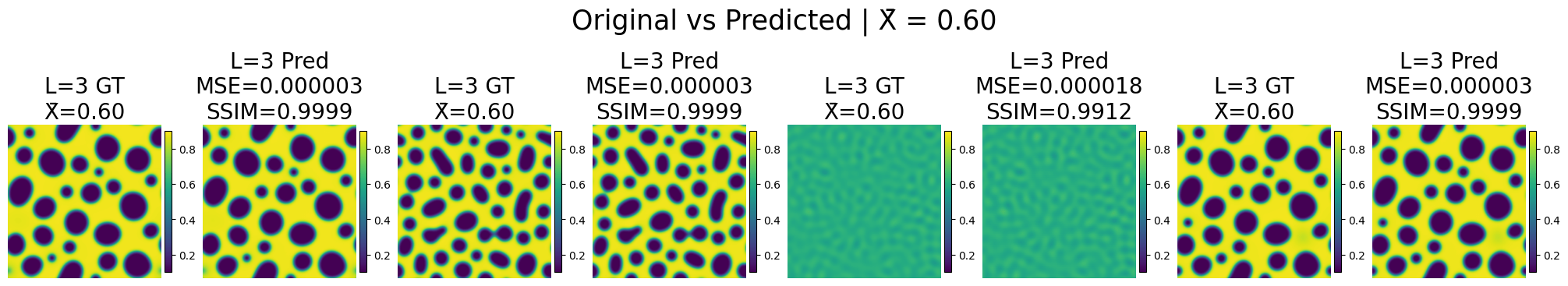}
\caption{GT and predicted microstructures for $\bar{X}_{\mathrm{Sb}}=0.5$ (top) and 
$0.6$ (bottom) at $L=3$. The GCN-LSTM-PI predicts the next-step latent state, 
decoded by the autoencoder, with results shown alongside GT, mean composition, 
MSE, and SSIM, highlighting reconstruction accuracy and conservation for the 
2D $256 \times 256$ model.}
\label{fig:ae_recon_combined}
\end{figure}
\begin{figure}[H] % needs \usepackage{float}
\centering
\includegraphics[width=0.50\textwidth]{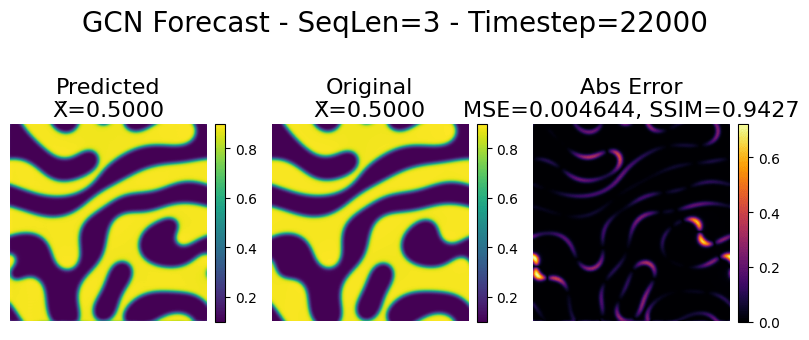}\\[0.5em]
\includegraphics[width=0.50\textwidth]{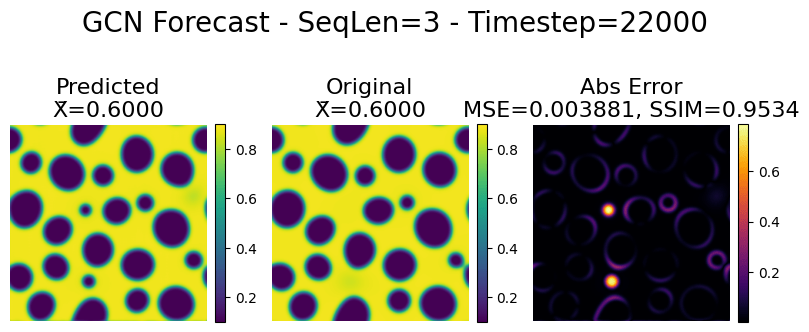}
\caption{GCN-LSTM-PI forecast results at $\bar{X}_{\mathrm{Sb}}=0.5$ (top), and $\bar{X}_{\mathrm{Sb}}=0.6$ (bottom) for the 2D $256 \times 256$ model. Each row shows the predicted microstructure, ground truth, and error map for future timesteps beyond the training
range. The forecasts display the predicted microstructures at 2,000 timesteps after the last training point.}
\label{fig:future_forecast_gcn}
\end{figure}

\begin{table}[h]
\caption{Long-horizon forecast metrics for sequence length $L = 3$ at future timesteps 
for $\mathbf{\bar{X}_{\mathrm{Sb}}} = 0.5$ and $0.6$ of the 2D model with input size $256\times256$}
\label{tab:forecast_metrics}
\begin{tabular*}{\textwidth}{@{\extracolsep\fill}lcccccccc}
\toprule
& \multicolumn{4}{@{}c@{}}{$\mathbf{\bar{X}_{\mathrm{Sb}}} = \mathbf{0.5}$} & \multicolumn{4}{@{}c@{}}{$\mathbf{\bar{X}_{\mathrm{Sb}}} = \mathbf{0.6}$} \\
\cmidrule{2-5}\cmidrule{6-9}
\textbf{Timestep} & \textbf{MSE} & \textbf{SSIM} & \textbf{$\mathbf{\bar{X}_{\mathrm{pred}}}$} & \textbf{$\mathbf{\bar{X}_{\mathrm{true}}}$} & \textbf{MSE} & \textbf{SSIM} & \textbf{$\mathbf{\bar{X}_{\mathrm{pred}}}$} & \textbf{$\mathbf{\bar{X}_{\mathrm{true}}}$} \\
\midrule
21000 & 0.001765 & 0.9749 & 0.5000 & 0.5000 & 0.000808 & 0.9854 & 0.6000 & 0.6000 \\
23000 & 0.007909 & 0.9099 & 0.5000 & 0.5000 & 0.005846 & 0.9310 & 0.6000 & 0.6000 \\
25000 & 0.015151 & 0.8496 & 0.5000 & 0.5000 & 0.012458 & 0.8841 & 0.6000 & 0.6000 \\
27000 & 0.023103 & 0.7982 & 0.5000 & 0.5000 & 0.016416 & 0.8483 & 0.6000 & 0.6000 \\
29000 & 0.031557 & 0.7553 & 0.5000 & 0.5000 & 0.020384 & 0.8162 & 0.6000 & 0.6000 \\
31000 & 0.040743 & 0.7177 & 0.5000 & 0.5000 & 0.026840 & 0.7845 & 0.6000 & 0.6000 \\
32000 & 0.045174 & 0.7009 & 0.5000 & 0.5000 & 0.029523 & 0.7706 & 0.6000 & 0.6000 \\
33000 & 0.049452 & 0.6853 & 0.5000 & 0.5000 & 0.032694 & 0.7579 & 0.6000 & 0.6000 \\
35000 & 0.057548 & 0.6572 & 0.5000 & 0.5000 & 0.035160 & 0.7419 & 0.6000 & 0.6000 \\
37000 & 0.065095 & 0.6322 & 0.5000 & 0.5000 & 0.039482 & 0.7224 & 0.6000 & 0.6000 \\
39000 & 0.072205 & 0.6099 & 0.5000 & 0.5000 & 0.043029 & 0.7055 & 0.6000 & 0.6000 \\
\botrule
\end{tabular*}
\end{table}

The last results discussed are based on the 3D architecture with an input size of 128$\times$128$\times$128. The training method for this model replicates the 128×128 setup and differs only in loss weights, with more emphasis on physics, conservation, and SSIM, and less on the latent term (the image MSE remains unchanged). The training hyperparameters are:

\begin{itemize}
  \item Epochs: 100
  \item Learning rate: \(1\times 10^{-4}\)
  \item Image MSE loss weight: \(\lambda_{\text{img}} = 2.0\)
  \item SSIM loss weight: \(\lambda_{\text{ssim}} = 2.0\)
  \item Conservation loss weight: \(\lambda_{\text{cons}} = 0.05\)
  \item Physics-informed loss weight: \(\lambda_{\text{phys}} = 3.0\)
  \item Latent loss weight: \(\lambda_{\text{latent}} = 0.05\)
\end{itemize}

\begin{figure}[H]
\centering
\includegraphics[width=0.9\textwidth]{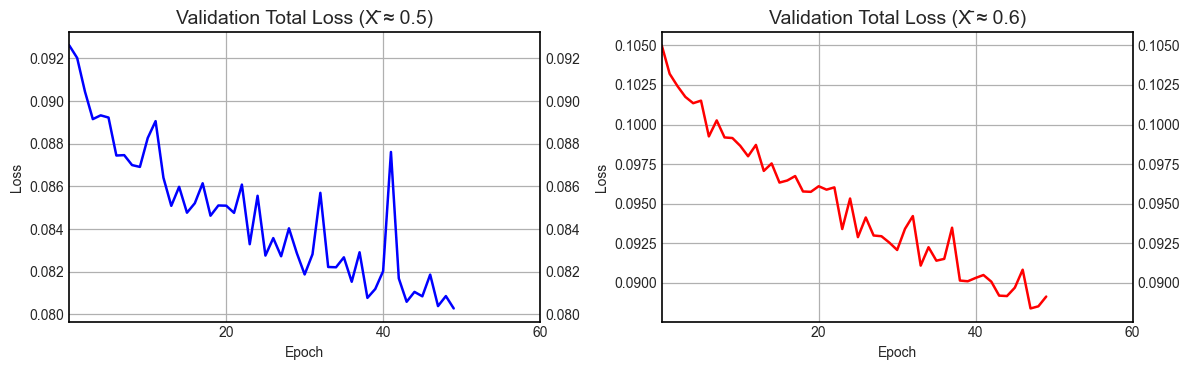}
\caption{Validation total loss over 100 epochs for the 3D GCN-LSTM-PI model 
($128^3$, $L=3$) at $\bar{X}_{\mathrm{Sb}}=0.5$ (left) and $0.6$ (right). 
The loss, combining SSIM, MSE, physics, conservation, and latent terms, 
decreases gradually but does not fully converge, reflecting model complexity 
and local computational limits.}
\label{fig5}
\end{figure}

\begin{table}[h]
\centering
\caption{Validation losses at epoch 70 of \textbf{$\bar{X}_{\mathrm{Sb}} = 0.5$, $\bar{X}_{\mathrm{Sb}} = 0.6$} for the 3D
GCN-LSTM-PI model with input size $128\times128\times128$}
\label{tab:x_losses_combined}
\begin{tabular*}{\textwidth}{@{\extracolsep\fill}lcccccc}
\toprule
\textbf{Compos.} & \textbf{Total Loss} & \textbf{MSE} & \textbf{Physics Loss} & \textbf{Latent Loss} & \textbf{Cons.\ Loss} & \textbf{SSIM} \\
\midrule
\(\bar{X}_{\mathrm{Sb}}=0.5\) & 0.900473 & 0.00268487 & 0.00728084 & 0.285801 & 0.0000000000 & 0.9600 \\
\(\bar{X}_{\mathrm{Sb}}=0.6\) & 0.775819 & 0.00351889 & 0.00745764 & 0.239749 & 0.0000000000 & 0.9473 \\
\botrule
\end{tabular*}
\end{table}

\begin{figure}[H]
\centering
\includegraphics[width=1\textwidth]{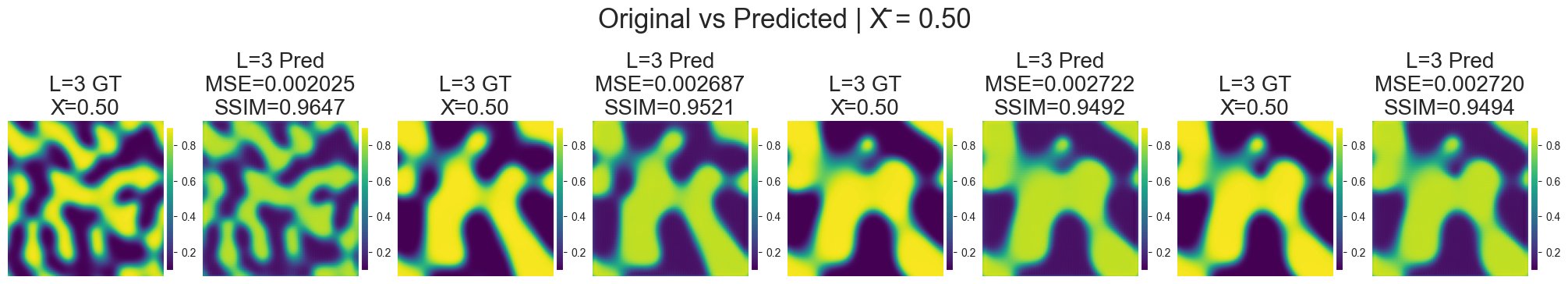}
\includegraphics[width=1\textwidth]{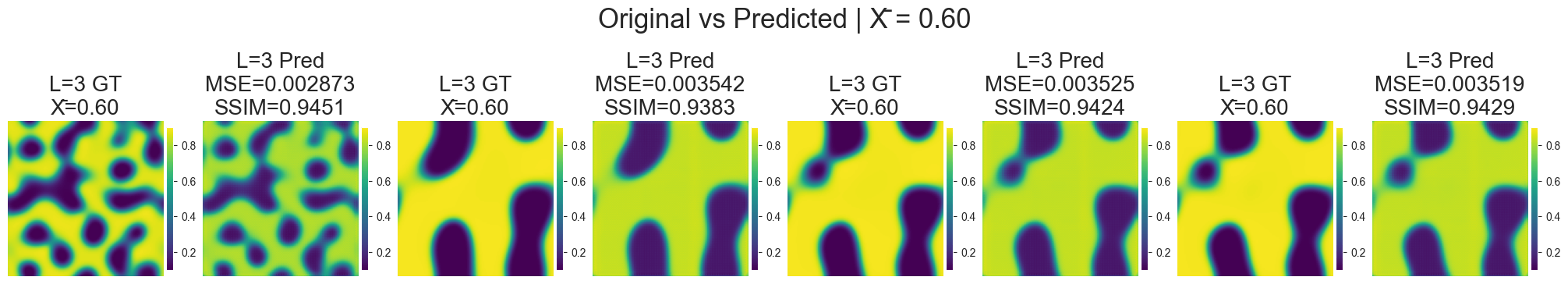}
\caption{Comparison of original (GT) and predicted microstructure images for 
$\bar{X}_{\mathrm{Sb}} = 0.5$ (top) and $\bar{X}_{\mathrm{Sb}} = 0.6$ (bottom) 
across sequence lengths $L = 3$. The GCN-LSTM-PI predicts 
the latent representation of the next timestep, which is decoded into a microstructure 
image using the trained autoencoder. Predictions are shown side by side with their 
ground truth, along with the corresponding mean composition, MSE, and SSIM values. 
This visualization highlights reconstruction accuracy and conservation across 
different input sequence lengths for the 3D model with input size $128 \times 128\times128$.}
\label{fig:ae_recon_combined}
\end{figure}

In a single window in Figure 26, the one-step 3D forecasts \(t = 22{,}000\) are displayed alongside the ground truth for two mean compositions for \(\bar{X}_{\mathrm{Sb}} = 0.50\) and \(\bar{X}_{\mathrm{Sb}} = 0.60\). Each row contains three panels: the Predicted, Ground Truth, and Absolute Error. The error panel displays the voxel-wise magnitude \(\lvert \hat{x} - x \rvert\), and brighter regions denote larger local discrepancies, while darker regions indicate close similarity.

\begin{figure}[H]
\centering
\begin{minipage}[b]{0.48\textwidth}
    \centering
    \includegraphics[width=\linewidth]{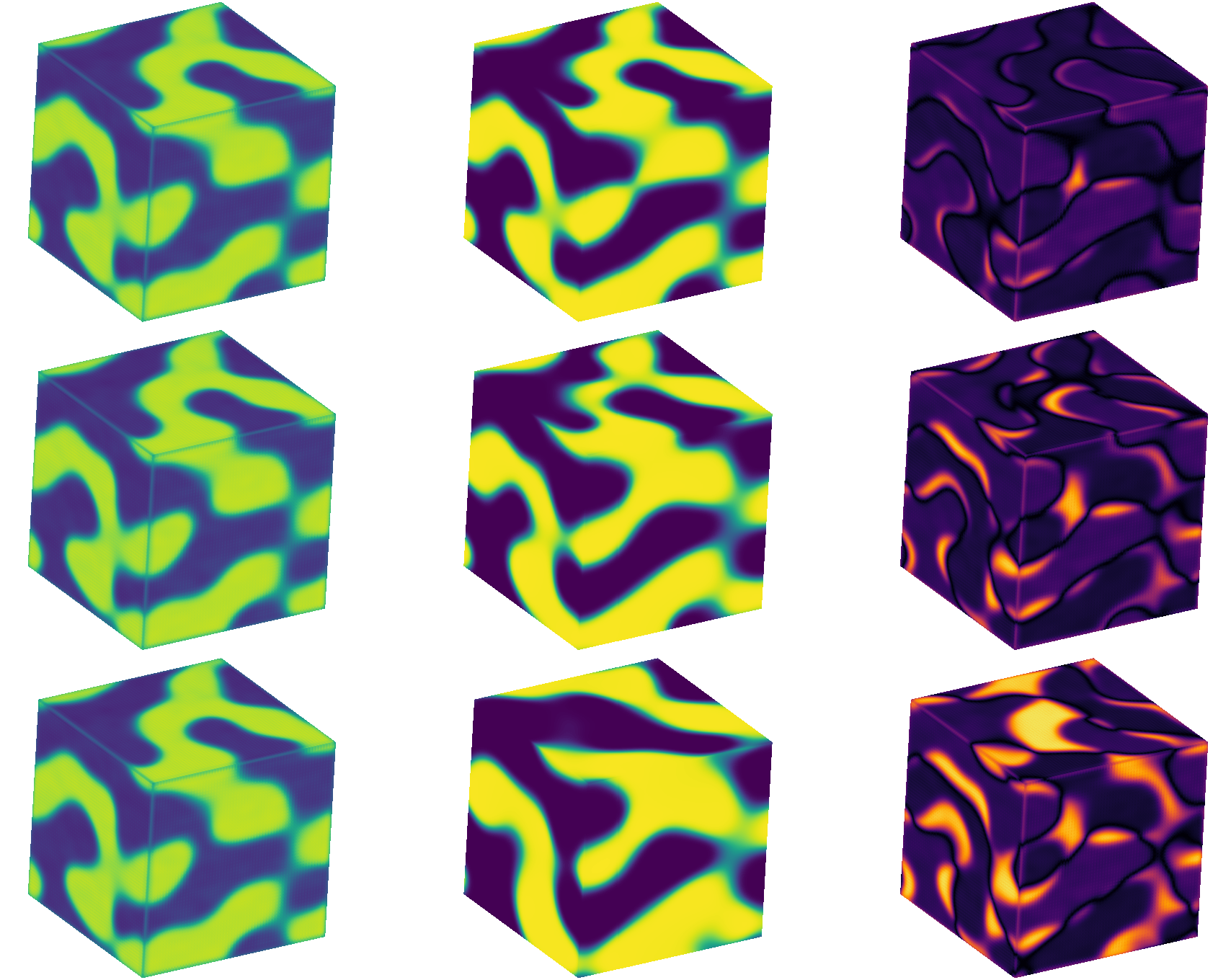}
\end{minipage}%
\hfill
\begin{minipage}[b]{0.48\textwidth}
    \centering
    \includegraphics[width=\linewidth]{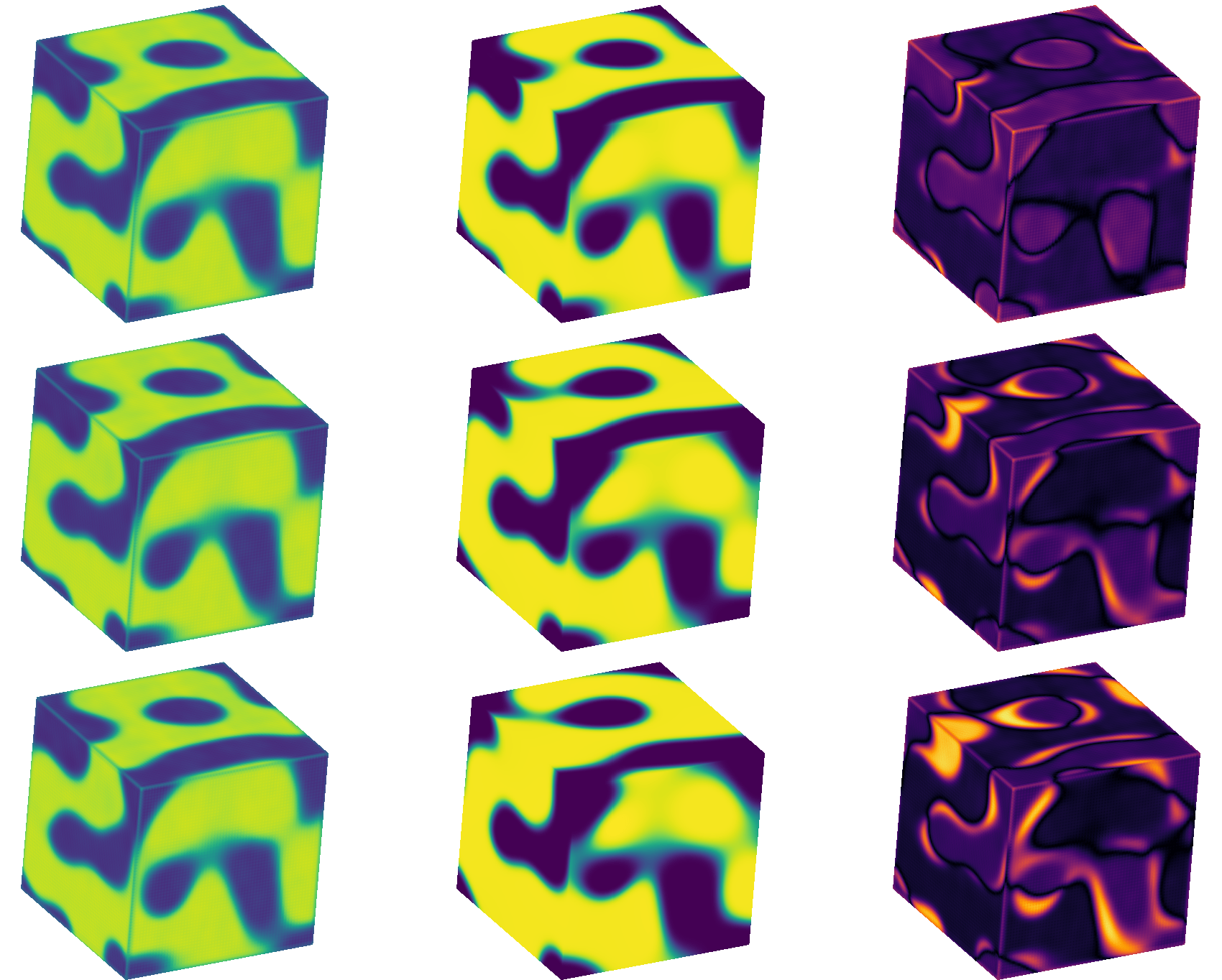}
\end{minipage}

 \caption{Long-horizon forecasting for the 3D model. The training and evaluation were performed on timesteps $t=0$ to $t=20{,}000$, and forecasting was carried out from $t=21{,}000$ to $t=39{,}000$. 3D forecasts at $t=23{,}000$ (top), $t=31{,}000$ (middle), and $t=39{,}000$ (bottom) versus ground truth for $\bar{X}_{\mathrm{Sb}}=0.50$ (left side) and $\bar{X}_{\mathrm{Sb}}=0.60$ (right side). Columns show the predicted microstructure (left side), the ground truth (middle images), and the absolute error volume $\lvert \hat{x}-x\rvert$ (right side). Brighter regions in the error panel indicate larger local discrepancies, whereas darker regions indicate closer agreement.}
\label{fig:combined}
\end{figure}
\begin{table}[h]
\centering
\caption{Long-horizon forecast metrics for sequence length $L = 3$ at future timesteps for $\bar{X}_{\mathrm{Sb}} = 0.5$ and $0.6$  of the 3D
GCN-LSTM-PI model with input size $128\times128\times128$.}
\label{tab:forecast_metrics}
\begin{tabular*}{\textwidth}{@{\extracolsep\fill}lcccccccc}
\toprule
& \multicolumn{4}{@{}c@{}}{$\mathbf{\bar{X}_{\mathrm{Sb}}} = \mathbf{0.5}$} & \multicolumn{4}{@{}c@{}}{$\mathbf{\bar{X}_{\mathrm{Sb}}} = \mathbf{0.6}$} \\
\cmidrule(lr){2-5}\cmidrule(lr){6-9}
\textbf{Timestep} & \textbf{MSE} & \textbf{SSIM} & \textbf{$\mathbf{\bar{X}_{\mathrm{pred}}}$} & \textbf{$\mathbf{\bar{X}_{\mathrm{true}}}$} & \textbf{MSE} & \textbf{SSIM} & \textbf{$\mathbf{\bar{X}_{\mathrm{pred}}}$} & \textbf{$\mathbf{\bar{X}_{\mathrm{true}}}$} \\
\midrule
21000 & 0.003417 & 0.9613 & 0.5000 & 0.5000 & 0.003455 & 0.9617 & 0.6000 & 0.6000 \\
23000 & 0.009690 & 0.8938 & 0.5000 & 0.5000 & 0.006262 & 0.9339 & 0.6000 & 0.6000 \\
25000 & 0.018714 & 0.8174 & 0.5000 & 0.5000 & 0.010341 & 0.9037 & 0.6000 & 0.6000 \\
27000 & 0.028707 & 0.7546 & 0.5000 & 0.5000 & 0.014951 & 0.8748 & 0.6000 & 0.6000 \\
29000 & 0.039452 & 0.7051 & 0.5000 & 0.5000 & 0.019838 & 0.8473 & 0.6000 & 0.6000 \\
31000 & 0.050773 & 0.6684 & 0.5000 & 0.5000 & 0.024917 & 0.8215 & 0.6000 & 0.6000 \\
32000 & 0.056387 & 0.6536 & 0.5000 & 0.5000 & 0.027518 & 0.8093 & 0.6000 & 0.6000 \\
33000 & 0.062061 & 0.6407 & 0.5000 & 0.5000 & 0.030158 & 0.7975 & 0.6000 & 0.6000 \\
35000 & 0.074527 & 0.6167 & 0.5000 & 0.5000 & 0.035548 & 0.7756 & 0.6000 & 0.6000 \\
37000 & 0.086881 & 0.5929 & 0.5000 & 0.5000 & 0.041101 & 0.7555 & 0.6000 & 0.6000 \\
39000 & 0.098066 & 0.5678 & 0.5000 & 0.5000 & 0.047003 & 0.7370 & 0.6000 & 0.6000 \\
\botrule
\end{tabular*}
\end{table}

\section{Conclusion}\label{sec6}

In conclusion, the introduced physics-informed framework integrates graph convolutional networks (GCN) with long-term short-term memory (LSTM) to forecast microstructure evolution over long time horizons in both 2D and 3D, achieving remarkable performance across various metrics. Encoding phase-field simulation data with convolutional autoencoders and operating in latent graph space enables the model to capture the compositions and morphological dynamics while remaining computationally efficient. The defined physics-informed losses based on the Cahn-Hilliard formulation and composition conservation allow physical consistency with the phase-field throughout the long-range predictions. By training the model jointly on multiple compositions, the framework can generalize its forecast across compositions, which facilitates rapid exploration of structural relationships beyond the training window. This approach offers a practical surrogate for accelerating phase-field studies, as it predicts long-term future states substantially faster than conventional simulations once trained. The ability to forecast across compositions, dimensions, and long-term horizons positions this method as a valuable approach in computational materials engineering. 

Future work will involve training the model on more varied compositions, improving future forecasting performance, additional physics (e.g., elasticity, anisotropy, temperature-dependent mobility), and implementing multi-component alloy systems. 

\section{Acknowledgment}\label{sec13}

We acknowledge the financial support from the European Research Council (ERC) under the European Union’s Horizon 2020 research and innovation program: grant agreement number 101123107 mTWIN -- Innovative digital twin concept of complex microstructure evolution in multi-component materials

\section*{Conflict of Interest}

The authors declare that they have no conflict of interest. 

\section*{Data and Code Availability}

Data and codes would be provided by the corresponding author on a reasonable request. 




%\end{appendices}


\bibliography{sn-bibliography}% common bib file

\end{document}